\documentclass[onecolumn,showpacs, superscriptaddress,preprint,nofootinbib,12pt]{revtex4-2}
\usepackage[pdftex]{graphicx}
\usepackage[pdftex]{epsfig}
\usepackage{subfigure}
\usepackage[titletoc,title]{appendix}
\usepackage{amssymb,amsmath,mathtools,dsfont}
\usepackage[hypertexnames=false,colorlinks,urlcolor=blue,citecolor=blue]{hyperref}
\usepackage[dvipsnames]{xcolor}
\usepackage{tikz}
\usetikzlibrary{quantikz}
\usepackage{cleveref}
\usepackage{soul}
\Crefname{section}{Section}{Sections}
\crefname{section}{Sec.}{Secs.}
\crefname{figure}{Fig.}{Figs.}
\crefname{table}{Table}{Tables}
\usepackage{{booktabs}}
\usepackage{setspace}
\usetikzlibrary{matrix,arrows,decorations.pathmorphing}
\newenvironment{sloppypar*}{\sloppy\ignorespaces}{\par}
\allowdisplaybreaks 

\newcommand {\sket} [1] {| #1 \rangle}

\newcommand\spec{\operatorname{spec}}

\newcommand {\unit} {\mathds{1}}

\newcommand{\pd}{{\vphantom{\dagger}}}

\newcommand{\T}{\mathrm{T}}

\begin{document}

\title{Exponential improvements in the simulation of lattice gauge theories using near-optimal techniques}   

\author{Mason L. \surname{Rhodes}}
\affiliation{Center for Computing Research, Sandia National Laboratories, Albuquerque, NM, 87185, USA}
\affiliation{Center for Quantum Information and Control, University of New Mexico, Albuquerque, NM, 87131, USA}
\affiliation{Department of Physics and Astronomy, University of New Mexico, Albuquerque, NM, 87131, USA}

\author{Michael Kreshchuk}
\affiliation{Physics Division, Lawrence Berkeley National Laboratory, Berkeley, California 94720, USA}
\affiliation{
Phasecraft Inc., Washington, D.C., 20001, USA
}

\author{Shivesh Pathak}
\affiliation{Quantum Algorithms and Applications Collaboratory, Sandia National Laboratories, Albuquerque, NM, USA}

\begin{abstract}
{We report a first-of-its-kind analysis on post-Trotter simulation of U(1), SU(2) and SU(3) lattice gauge theories including fermions in arbitrary spatial dimension.
We provide explicit circuit constructions as well as $\T$-gate counts and logical qubit counts for Hamiltonian simulation.} 
{We find up to 25 orders of magnitude reduction in spacetime volume over Trotter methods for simulations of non-Abelian lattice gauge theories relevant to the standard model. 
This improvement results from our algorithm having polynomial scaling with the number of colors in the gauge theory, achieved by utilizing oracle constructions relying on the sparsity of physical operators, in contrast to the exponential scaling seen in state-of-the-art Trotter methods which employ explicit mappings onto Pauli operators.
Our work demonstrates that the use of advanced algorithmic techniques leads to dramatic reductions in the cost of simulating fundamental interactions, bringing it in step with resources required for first principles quantum simulation of chemistry.}
\end{abstract}

\maketitle

\tableofcontents
\singlespacing 
\section{Introduction \label{sec:1}}
    
    The simulation of strongly interacting quantum many-body systems has continues to be a challenging problem in classical and quantum simulation.
    Simulations of quantum chemistry \cite{C002859B}, condensed matter physics \cite{kononov2023}, and lattice quantum chromodynamics~\cite{DeTar:2004a,Nagata:2022a} consume a significant portion of the world's high-performance computing resources.
    The difficulty of accurately and efficiently performing such simulations on classical computers inspired the development of the field of quantum computing~\cite{feynman1982}, and is still considered one of its most exciting potential applications~\cite{Bauer:2023a,Bauer:2023c,Fauseweh:2024a}.
    Substantial effort has already been devoted to efficiently simulating electronic structure from first principles, employing sophisticated algorithms based on both first- and second-quantized formulations~\cite{Aspuru-Guzik:2005a,Whitfield:2011a,Hastings:2014a,Berry:2014a,Wecker:2014a,McClean:2014a,Poulin:2014a,Babbush:2015a,Babbush:2016a,Babbush:2018a,Babbush:2018b,Kivlichan:2018a,Berry:2018a,Babbush:2019a,Berry:2019a,Kivlichan:2020a,Bauer:2020a,Lee:2021a,Su:2021b,Su:2021a,Goings:2022a,Rubin:2023b,Berry:2023a,Loaiza:2023a,Rubin:2023a,Watson:2023oov,Du:2023bpw,Liu:2024hmm,Du:2024zvr,Wang:2024scd, pathak2023, pathak2024}.
    Owing to its classical hardness~\cite{Jordan2018bqpcompletenessof}, of particular interest, is utilizing quantum computers for dynamical simulations of fundamental interactions, particularly non-Abelian lattice gauge theories (LGTs).
    
    {There is a significant corpus of work on the quantum simulation of the dynamics of LGTs using Trotter product formula techniques. 
    Studies include simulation algorithms of Abelian and non-Abelian finite gauge groups}~\cite{Gustafson:2022a,Gustafson:2023a,Gustafson:2024a,Lamm:2024a,Assi:2024a, Zohar:2017a,Bender:2018a,Lamm:2019a, Carena:2021a, Carena:2022a, pardo2023resource} {as well as Lie groups} \cite{Davoudi:2022xmb, Byrnes:2006a, Mathis:2020a,Shaw:2020a, Ciavarella:2021a}, {from 1 to 3 spatial dimensions} \cite{Mathis:2020a,Shaw:2020a,Carena:2021a,Ciavarella:2021a, Byrnes:2006a,Zohar:2017a,Bender:2018a,Lamm:2019a}, {constructed for both noisy intermediate-scale quantum (NISQ) hardware and fault-tolerant hardware (FT), and in some cases with rigorous resource estimates for NISQ and FT regimes}~\cite{Davoudi:2022xmb, Mathis:2020a,Shaw:2020a} {or even experimental realizations on contemporary NISQ hardware}~\cite{Carena:2021a, Ciavarella:2021a, Carena:2022a, pardo2023resource}.
    {While these papers constitute a thorough literature in tandem, each result is limited, such as only including implementations for a particular gauge group, being limited to a particular spatial dimension, or only simulating pure gauge theories without fermions.
    One standout paper, however, by Kan and Nam}~\cite{kan2021} {provides a quantum algorithm and FT resource estimates for Trotterized simulation of lattice gauge theories applicable for 1 to 3 spatial dimensions, including fermions, and for U(1), SU(2) and SU(3) LGTs, and has been a benchmark of comparison in the quantum simulation of LGTs.
    }

    Trotter-based methods, while having simple implementations and prospectively small constant-factor overheads \cite{PhysRevX.11.011020}, have suboptimal asymptotic scaling in parameters such as problem size, evolution time, gauge group dimension, spatial dimension, and error.
    A particular challenge in LGT simulation is the exponential scaling in the number of color degrees of freedom in state-of-the-art Trotter methods \cite{Davoudi:2022xmb, kan2021}, resulting in dramatic increases in gate counts from $10^{23}$ expensive non-Clifford operations for FT simulation of U(1) in 2+1D, to $10^{40}$ for SU(2) and $10^{55}$ for SU(3) gauge theories \cite{kan2021}.
    As such, simulation of LGTs using post-Trotter methods that possess optimal or near-optimal dependence problem parameters ~\cite{Tong:2022a, low2017optimal,Low:2019a,Gily_n_2019,Lin:2020zni,Lin:2022a,bebyqsp} --- which we collectively refer to as \emph{qubitization} (see~\Cref{app:qubitization}) --- may be advantageous over Trotter methods in certain parameter regimes.
    However, to date, no comprehensive study of qubitized simulation of LGTs has been carried out to address these potential advantages, with 
    the few extant results relegated to small system sizes using the quantum eigenvalue transformation~\cite{Kane:2023jdo}, {light-front quantization}~\cite{Kreshchuk:2020dla,kirby2021quantum}, {and quantum signal processing techniques}~\cite{Tong:2022a,Hariprakash:2023tla,Anderson:2024kfj,kane2024block}.

    {In this work, we build on and fully develop the algorithm for qubitized Hamiltonian simulation of LGTs, based on the proposal by Tong \textit{et al.} \cite{Tong:2022a}, and work out all the necessary details for U(1), SU(2), and SU(3) gauge theories coupled to fermions in arbitrary spatial dimensions.}
    We lay out the ground work of geometrically local fermion-to-qubit encodings and construct explicit block-encoding oracle circuits using the linear combination of unitaries (LCU) and sparse oracle constructions.
    We also estimate all relevant algorithmic parameters, including Lieb-Robinson velocities, and provide complete front-to-back FT resource estimates in terms of T-gate counts and logical qubit counts for qubitized Hamiltonian simulation.
     
    Importantly, when comparing our results to Kan and Nam's work on Trotterized simulation~\cite{kan2021}, we find up to 25 orders of magnitude improvements in $\T$-gate counts for SU(3) LGT simulation.
    This is primarily due to a polynomial scaling in color degrees of freedom in the qubitized simulation compared to exponential scaling in Trotterized simulation.
    {Such efficiency was achieved by utilizing oracle constructions relying on the sparsity of physical operators rather than on their explicit mappings onto Pauli operators.}
    {We note that while Kan and Nam's work is not representative of the most optimal Trotterized simulation algorithm in terms of gate counts, as the aforementioned corpus of work contains optimizations for specific problem instances \cite{Davoudi:2022xmb, Byrnes:2006a, Mathis:2020a,Shaw:2020a, Ciavarella:2021a, Mathis:2020a,Shaw:2020a,Carena:2021a,Ciavarella:2021a, Byrnes:2006a,Zohar:2017a,Bender:2018a,Lamm:2019a}, {our work is, to the best of our knowledge, the first to have polynomial scaling in color degrees of freedom.}}

    The contents of the paper are as follows.
    In~\cref{sec:2} we discuss Hamiltonian discretization using the Kogut-Susskind formalism, followed by explicit constructions for geometric localization of LGTs by means of various fermion-to-qubit mappings in~\cref{sec:3}.
    In~\cref{sec:4} we provide explicit constructions of circuits for sparse and LCU block encodings of U(1) gauge theory and comparison of the full interaction picture qubitized simulation to Trotterization: we also provide two unique constructions, an optimized Signed Increment-Decrement circuit, and bounding of Lieb-Robinson velocities for U(1) in dimension $d > 1$.
    In~\cref{sec:5} we provide generalizations of the constructions to SU(2) and SU(3) gauge theories, and conclude with a discussion of outlooks for simulation of LGTs in~\cref{sec:6}.

\section{Hamiltonian simulation \label{sec:2}}
\subsection{Kogut-Susskind Hamiltonian\label{sec:ks}}
Throughout this paper, we will follow the notations used by Kan and Nam \cite{kan2021} {for the Kogut-Susskind Hamiltonian}.
The Hamiltonian is defined over a $d$-dimensional lattice where each site is denoted by a vector $\vec{n} \in \mathbb{Z}^d$.
The lattice sites are labeled even or odd, depending on the sign $(-1)^{\vec{n}}= (-1)^{\sum_i n_i}.$
We will take the lattice to be a hypercube with linear dimension $N$ and total $N^d$ sites.

Fermions are distributed on the lattice sites, with the spinor being staggered such that the fermionic and anti-fermionic spinors are on even and odd sites, respectively.
Staggering fermions reduces the number of spurious fermionic degrees of freedom known as \emph{doublers}, which arise due to the periodic nature of the dispersion relation of lattice fermions~\cite{kogutsusskind1975,zohar2015formulation}.
For gauge theories with various fermion colors, each site contains multiple spinors corresponding to each color degree of freedom.
A link between neighboring sites is denoted by the tuple $(\vec{n}, \hat{l})$ where $\hat{l}$ indicates the direction of a nearest neighbor site to $\vec{n}$.
The links contain the gauge degrees of freedom which mediate the interaction between the fermionic degrees of freedom.

The Hamiltonian itself contains four terms:\footnote{Following the literature on quantum simulation~\cite{kan2021,Tong:2022a}, we use simplified notations for the Kogut-Susskind Hamiltonian. Complete expressions for the 2+1D and 3+1D cases can be found in Refs.~\cite{susskind1977lattice,crippa2024towards}.
}
\begin{equation}
    H = g_M H_M + g_{GM} H_{GM} + g_E H_E + g_B H_B\,.
    \label{eq:KS_full_ham}
\end{equation}
Here the four constants are related to the physical parameters, namely the dimension $d$, the lattice spacing $a$, the bare gauge-matter coupling constant $g$, and the bare fermion mass $m$ as:
\begin{equation}
    g_M = m\,, \ g_{GM}
    =
    \frac{1}{2a}\,, \ g_{E} = \frac{g^2}{2a^{d-2}}\,, \
    g_{B} = -\frac{1}{2a^{4-d}g^2}\,.
\end{equation}

The first of the four operators is the mass term:
\begin{equation}
    H_M = \sum_{\vec{n}}\sum_a (-1)^{\vec{n}} \psi_a^\dagger(\vec{n}) \psi_a^\pd(\vec{n})\,.
    \label{eq:KS_mass}
\end{equation}
Here $\psi_a^\pd(\vec{n})$ and $\psi_a^\dagger(\vec{n})$ are the creation and annihilation operators for a fermion with color degree of freedom $a$ on site $\vec{n}$.
The $(-1)^{\vec{n}}$ sign is indicative of the staggered nature of the fermions.
The next three terms are the gauge-matter, electric field, and magnetic field terms:
\begin{align}
    &H_{GM} = \sum_{\vec{n}, \hat{l}} \sum_{a, b} \psi_a^\dagger(\vec{n}) U_{ab}^\pd(\vec{n}, \hat{l})\psi_b^\pd(\vec{n} + \hat{l}) + \text{h.c.}\,,\label{eq:KS_gauge_matter} \\
    &H_{E} =  \sum_{\vec{n}, \hat{l}} \sum_a [E_a(\vec{n}, \hat{l})]^2\,, \label{eq:KS_electric} \\
    &H_B = \sum_{\square}\text{Tr}[P_\Box^\pd + P^\dagger_\Box]\,. \label{eq:KS_magnetic}
\end{align}
The plaquette operator $P_\Box^\pd$ is:
\begin{equation}
    \text{Tr}[P_\Box^\pd] = \sum_{a,b,c,e} U_{ab}^\pd(\vec{n}, \hat{i})U_{bc}^\pd(\vec{n}+ \hat{i}, \hat{j})U^\dagger_{ce}(\vec{n} + \hat{j}, \hat{i})U^\dagger_{ea}(\vec{n},\hat{j})\,,
    \label{eq:plaquette_term}
\end{equation}
where $\hat{i}, \hat{j}$ are links which form a closed loop starting and ending at a site $\vec{n}.$

The gauge link operator $U_{ab}(\vec{n}, \hat{l})$ and electric field operator $E_b(\vec{n}, \hat{l})^2$ depend on the gauge field under consideration.
For U(1) theory, the color index can be suppressed.
In this case, we further note that the Casimir operator $E^2$ and the electric field operator $E$ have the same eigenfunctions, which we label as $|k\rangle$ for $k \in \mathbb{Z}$. 
The operators act on these basis states as follows:

\begin{align}
    &E|k\rangle = k|k\rangle\,, \label{eq:electric_field_eigenstate}\\ 
    &U|k\rangle = |k-1\rangle\,, \label{eq:gauge_operator}
\end{align}    
i.e. $E$ and $U$, $U^\dagger$ act as the number and ladder operators, correspondingly.\footnote{Electric field digitization in Eq.~\eqref{eq:electric_field_eigenstate} is analogous to digitization of scalar fields considered in Refs.~\cite{2011arXiv1112.4833J,klco2018digitization, Macridin:2018a,Farrelly:2020a,Li:2022a,Hanada:2022a}.}
For a practical calculation, the number of bosonic modes is truncated such that $k\in [-\Lambda, \Lambda]$ where $\Lambda \in \mathbb{Z}^+$ is some integer.

For the case of the non-Abelian SU(2) and SU(3) theories, the situation is more complicated.
Each $E_a$ can no longer be simultaneously diagonalized as they do not commute, and instead one operates within the eigenbasis of the Casimir operator. 
For SU(2), the action of bosonic operators in this basis is given by
\begin{align}
&E^2|j, m^L, m^R\rangle = j(j+1)|j, m^L, m^R \rangle\,,\label{eq:su2-electric} \\
&\begin{aligned}
U_{ab}|j, m^L, m^R\rangle &= \sum_{J=|j - 1/2|}^{j + 1/2} \sqrt{\frac{2j+1}{2J+1}} \langle J,M_L|j, m^L; 1/2, a^\prime \rangle \langle J, M_R | j, m^R; 1/2,b^\prime \rangle \\
    &\times |J, M_L = m^L + a^\prime, M_R = m^R + b^\prime \rangle.  \label{eq:su2-magnetic}  
\end{aligned}
\end{align}
The first two terms including $J, M_L$ and $J, M_R$ in the operation of $U_{ab}$ are the Clebsch-Gordan coefficients for SU(2). Conservation of angular momentum along the $z$-axis requires that $a',b'=-1/2 \text{ or } 1/2$ when $a,b=1\text{ or }2$, respectively.
For SU(3) a similar construction applies, with the link basis again being the eigenbasis of the SU(3) Casimir operator,
\begin{align}
&E^2|p,q,T_L^{\phantom{z}},T_L^z,Y_L^{\phantom{z}},T_R^{\phantom{z}},T_R^z,Y_R^{\phantom{z}}\rangle=\frac{1}{3}\left(p^2+q^2+pq+3(p+q)\right)|p,q,T_L^{\phantom{z}},T_L^z,Y_L^{\phantom{z}},T_R^{\phantom{z}},T_R^z,Y_R^{\phantom{z}}\rangle\,,\label{eq:su3_electric} \\
&\begin{aligned}
&U_{ab}|p,q,T_L^{\phantom{z}},T_L^z,Y_L^{\phantom{z}},T_R^{\phantom{z}},T_R^z,Y_R^{\phantom{z}}\rangle \\
    &=\sum_{(p',q')}\sum_{T_L'=|T_L-t_L|}^{T_L+t_L}\sum_{T_R'=|T_R-t_R|}^{T_R+t_R}\sqrt{\frac{\text{dim}(p,q)}{\text{dim}(p',q')}}\langle p',q',T_L',T_L^{z'},Y_L'|p,q,T_L^{\phantom{z}},T_L^z,Y_L^{\phantom{z}};1,0,t_L^{\phantom{z}},t_L^z,y_L^{\phantom{z}}\rangle \\  
    &\times \langle p',q',T_R',T_R^{z'},Y_R'|p,q,T_R^{\phantom{z}},T_R^z,Y_R^{\phantom{z}};1,0,t_R^{\phantom{z}},t_R^z,y_R^{\phantom{z}}\rangle |p',q',T_L',T_L^{z'},Y_L',T_R',T_R^{z'},Y_R'\rangle\,,\label{eq:su3_gauge}
\end{aligned}
\end{align}
where $p,q\in[0,\Lambda]$ are SU(3) representation labels, and $T$, $T^z$, and $Y$ are known as the isospin, $z$-axis projection of the isospin, and hypercharge, respectively. Additionally, we have $T\in[0,\Lambda]$ and $T^z\in[-\Lambda,\Lambda]$ where each are incremented in half-integer steps, and $Y\in[-\Lambda,\Lambda]$ incremented in third-integer steps. For the gauge operators, $(p',q')\in\{(p+1,q),(p-1,q+1),(p,q-1)\}$ and $\text{dim}(p,q)=(1+p)(1+q)(1+(p+q)/2)$. Here, the isospin and hypercharge values $t$, $t^z$, and $y$ depend on the color degrees of freedom, but ultimately correspond to incrementing or decrementing the values of $T$, $T^z$, and $Y$, respectively, by half-integer or third-integer amounts. Finally, $\langle p',q',T',T^{z'},Y'|p,q,T,T^z,Y;1,0,t,t^z,y\rangle$ are the Clebsch-Gordan coefficients for SU(3) in the fundamental representation. For more details on the SU(3) gauge operators see Ref.~\cite[Appendix D]{kan2021}.

\subsection{Trotterization\label{sec:trotterization}}
{Trotterized approaches to Hamiltonian simulation invoke a} product formula {approximating} the exact time evolution operator as
\begin{equation}
    e^{iH T}=\left(e^{iH T/r}\right)^r\approx\left[\left(\prod_{j=1}^{N_H}e^{iH_j T/2r}\right)\left(\prod_{j=N_H}^1 e^{iH_j T/2r}\right)\right]^r,
\end{equation}
where we have decomposed the full Hamiltonian into the $N_H=4$ subterms of Eq.~\eqref{eq:KS_full_ham}, i.e. $j\in\{M,GM,E,B\}$, and we have decomposed the evolution into $r\gg T$ Trotter steps.
A number of algorithms exist which explore Trotterized approaches to simulating dynamics of LGTs~\cite{Byrnes:2006a,Zohar:2017a,Bender:2018a, Shaw:2020a,Mathis:2020a,Ciavarella:2021a,Lamm:2019a,Harmalkar:2020a,kan2021,Carena:2021a,Carena:2022a,Mathis:2020a,Davoudi:2022xmb,pardo2023resource}.
We focus in particular on the algorithms presented in Ref.~\cite{kan2021} as it contains the most {general} estimates provided to date {including U(1), SU(2) and SU(3) LGTs in $d$ spatial dimensions using a second-order Suzuki-Trotter formula}~\cite{Suzuki:1991a}.
{We emphasize that while other works cited provide significantly improved algorithms to the work of Kan and Nam, explicit resource estimates are limited to particular gauge groups or spatial dimensions (such as SU(2) in 1+1D}~\cite{Davoudi:2022xmb}{) thereby lacking the generality necessary for comparison to our work.}

The number of Trotter steps contributes a multiplicative prefactor to the cost of the full simulation complexity, and individual operators $\exp(iH_j T/2r)$ can be realized using polylogarithmic-size circuits. 
In particular, when considering the complexity scaling with respect to the bosonic truncation parameter $\Lambda$, the second-order Suzuki-Trotter formula requires only $\mathcal{O}(\Lambda)$ Trotter steps~\cite{Shaw:2020a}. 
On the other hand, typical qubitization approaches have a complexity scaling proportional to the Hamiltonian norm, which is dominated by the electric field term since $\|H_E\|=\mathcal{O}(\Lambda^2)$ {in the electric field basis for the gauge field.}
Thus, the product formula approach seems to admit a quadratic speedup over typical qubitization approaches, but this overhead can be avoided by performing simulation in the interaction picture {or by moving into a different basis for the gauge fields} \cite{PhysRevD.109.074501, Lamm:2024a}. This will be discussed further in Sec.~\ref{sec:qubitization}. 
In addition to the bosonic truncation parameter, the number of Trotter steps will also scale with the total evolution time $T$ and the precision of the approximation $\epsilon$ as $\mathcal{O}(T^{3/2}/\epsilon^{1/2})$~\cite{Shaw:2020a} for the second order product formula.

In Ref.~\cite{kan2021}, a binary encoding is used for the bosons stored on each lattice edge, but the Jordan-Wigner transformation is used to encode the fermions {which does not maintain geometric locality of the resulting qubit Hamiltonian.}
They then construct explicit circuits for each of the operators $\exp(iH_j T/2r)$ with respect to these encodings for U(1), SU(2), and SU(3) LGTs. 
The full asymptotic gate complexities are
\begin{align}
    \text{U(1)}&\sim \mathcal{O}(2^{8}N^{3d/2}T^{3/2}\Lambda\epsilon^{-1/2}\text{polylog}(\Lambda))\,, \\
    \label{eq:su2su3cost}
    \text{SU(2), SU(3)}&\sim \mathcal{O}(2^{8(n_c^2-1)}N^{3d/2}T^{3/2}\Lambda\epsilon^{-1/2}\text{polylog}(N^{3d/2}T^{3/2}\Lambda\epsilon^{-3/2}))\,,
\end{align}
where we have explicitly included the prefactor scaling exponentially in the number of colors responsible for the several orders of magnitude increase in complexity when scaling up from U(1) to SU(2) and SU(3) in Ref.~\cite{kan2021}. 

This exponential prefactor is a result of the decomposition the authors perform to block-diagonalize the Hamiltonian terms in order to implement their Trotterized time evolution more efficiently. The decomposition splits the Hamiltonian into a sum over all possible parity configurations in which off-diagonal elements can be written as $D\cdot P$ where $D$ is a diagonal operator and $P$ is a Pauli string. {T}o obtain the sum over all local quantum numbers, $D\cdot P$ can be conjugated with the possible parity operations over sums involving color indices.

{Despite} the exponential prefactor, the authors of Ref.~\cite{kan2021} note that this is a superpolynomial improvement over the results of Ref.~\cite{Byrnes:2006a} which scale as $\mathcal{O}(\Lambda^{4(n_c^2-1)})$ for SU(2) and SU(3). However, we note that this is only a quadratic improvement over a na\"ive Pauli decomposition. Each gauge field operator of size $(n_c^2-1)\log\Lambda \times (n_c^2-1)\log\Lambda$ can be represented as a string of $\Lambda^{2(n_c^2-1)}$ Pauli operators. Then the magnetic field term, a product of four gauge field operators, will dominate, scaling as $\Lambda^{8(n_c^2-1)}$. We emphasize this prefactor here to compare to our algorithm later in Sec.~\ref{sec:qubitization}, {which} scales polynomially in $n_c$.

\subsection{Qubitization\label{sec:qubitization}}
While {various} qubitization-based approaches to simulating LGTs have been considered in the literature~\cite{Hariprakash:2023tla,Anderson:2024kfj}, we focus on an efficient algorithm proposed in Ref.~\cite{Tong:2022a} which has a near-optimal scaling in the space-time volume.
We consider a $d$-dimensional lattice with total number of sites $N^d$, for which we want to carry out dynamical simulation of some geometrically-local Hamiltonian $H$ for time $T$ with error $\epsilon$, starting from a state with local quantum numbers between $[-\Lambda_0, \Lambda_0]$ at any gauge link.

The core of the algorithm is the HHKL decomposition~\cite{hhkl2021}, which allows us to decompose full time evolution into time evolution of local blocks $\mathcal{B}$, of size $N_\mathcal{B}^d = \mathcal{O}(\text{polylog}(N^dT\epsilon^{-1}))$ for which we only require simulation for time $\tau = \mathcal{O}(1)$.
The entire simulation of the system then requires $\mathcal{O}(T)$ such segments and $\mathcal{O}(N^d)$ such blocks.
The HHKL decomposition is only valid under certain assumptions on the geometric locality of the Hamiltonian, all of which are satisfied by the LGTs under consideration here \cite{Tong:2022a}.

One can then consider the time evolution of the Hamiltonian restricted to the local blocks $\mathcal{B}$, which are polylogarithmic in the system size.
Rather than constructing $e^{itH^\mathcal{B}}$ directly, one finds that for LGTs using interaction picture dynamics yields a particularly efficient implementation.
Specifically, one can fast-forward the dynamics of the electric field term, $e^{itH^\mathcal{B}_E}$, as it is entirely diagonal in the link basis, regardless of the gauge theory.
This removes a quadratic dependence on the bosonic cutoff in the overall scaling of the algorithm that would otherwise be necessary to block encode the electric field term.
Since each of the electric field terms acts on separate links, all of these terms will commute and their simulation can be performed in parallel, so simulation of $e^{itH_E^\mathcal{B}}$ for each local block has only $\mathcal{O}(N_\mathcal{B}^d\text{polylog}(\Lambda_0 t)) = \mathcal{O}(\text{polylog}(N^dT\Lambda_0 t \epsilon^{-1}))$ gate complexity.

After fast-forwarding, one can simulate dynamics with the time-dependent interaction-picture Hamiltonian:
\begin{equation}
    H^\mathcal{B}_I(t) = e^{it g_E^{\phantom{\mathcal{B}}} H^\mathcal{B}_E}(g_M^{\phantom{\mathcal{B}}} H^\mathcal{B}_{M} + g_{GM}^{\phantom{\mathcal{B}}} H^\mathcal{B}_{GM} + g_B^{\phantom{\mathcal{B}}} H^\mathcal{B}_{B})e^{-it g_E^{\phantom{\mathcal{B}}} H^\mathcal{B}_E}\,.
    \label{eq:interaction-block}
\end{equation}
By making use of a truncated Dyson series implementation \cite{Low:2019a}, one can simulate dynamics of $H^\mathcal{B}_I(t)$ using $\mathcal{O}(\alpha t \text{polylog}(\alpha t\epsilon^{-1}))$ queries to the time-dependent block encoding of $H^\mathcal{B}_I(t)$, where $\alpha = \text{max}_{s \in (0, t)}\|H^\mathcal{B}_I(s)\|$, referred to as HAM-T. 
Since each term $H_M, H_{GM}, H_{B}$ is bounded by a constant, the total norm scales simply as $N_\mathcal{B}^d.$
Therefore, $\mathcal{O}(\text{polylog}(N^dT\Lambda_0 \epsilon^{-1}))$ queries to $H^{\mathcal{B}}_I(t)$ are required.
The implementation of $H^{\mathcal{B}}_I(t)$ itself must scale as $N_\mathcal{B}^d$, as there are only $N_\mathcal{B}^d$ terms in each  Hamiltonian.
Therefore, the final asymptotic gate cost should still be $\mathcal{O}(\text{polylog}(N^dT\Lambda_0 \epsilon^{-1}))$.
Finally, given that we must simulate $\mathcal{O}(N^d)$ such blocks at each of the $\mathcal{O}(T)$ time segments, the total asymptotic cost is:
\begin{equation}
    \text{Gate Cost} \sim \mathcal{O}(n_c^4 N^dT\text{polylog}(N^dT\Lambda_0 \epsilon^{-1}))\,,
\end{equation}
The qubitized algorithm implements the gauge operators directly, meaning for a single plaquette of the magnetic field term, we only must sum over the $n_c^4$ possible color configurations at the plaquette vertices. 
Thus, by using qubitization, we achieve exponential improvement in algorithm complexity with respect to the number of colors compared to the Trotterized approach of Ref.~\cite{kan2021} which has scaling $\mathcal{O}(2^{8(n_c^2-1)})$.
In this work, we will establish rigorous resource estimates for qubitized simulations of LGTs for various gauge theories and dimensions.

\section{Geometrically local qubit encodings \label{sec:3}}
As a final technical stepping stone to resource estimates, we provide explicit constructions for boson- and fermion-to-qubit encodings for arbitrary dimension $d$ and gauge theory SU(N).
Importantly, we provide constructions that preserve geometric locality of the Kogut-Susskind Hamiltonian, a necessary component of efficient simulation using the qubitized algorithm involving the HHKL decomposition as described in Sec.~\ref{sec:qubitization}.

\subsection{Boson-to-qubit mapping}
The Hilbert space of each bosonic Degree of Freedom (DOF) is infinite-dimensional, meaning a truncation on the space is necessary in order to achieve a tractable encoding. The tradeoff between the precision of the result and level of truncation is finely tuned, but a provable bound was established in Ref.~\cite{Tong:2022a}.
In particular, given an initial state in which the local quantum number on every link is within $[-\Lambda_0,\Lambda_0]$, evolving a truncated Hamiltonian for time $T$ on $N^d$ lattice sites to error $\epsilon$ can be achieved by keeping bosonic modes only in the range of $[-\Lambda, \Lambda]$, where
\begin{equation}
    \Lambda=\Lambda_0+\tilde{\mathcal{O}}((\chi T+1)\text{polylog}(N/\epsilon))\,,
\end{equation}
with $\chi$ being a constant depending solely on the model parameters and not $N$ and $T$. For LGTs, we have $\chi=c(4|g_B|+2|g_{GM}|)$ with $c$ being the number of (color indexed) gauge operators $U_{ab}$ in the theory.
This bound was further improved in Ref.~\cite{Peng:2023a} to
\begin{equation}
    \Lambda=\Lambda_0+\mathcal{O}\!\left(\chi T\log(N\Lambda_0\chi T/\epsilon)\right).
\end{equation}

There are two main approaches taken for direct boson-to-qubit encodings~\cite{sawaya2020resource}.
A unary encoding requires a linear number of qubits but fewer gates to implement oracles while a binary encoding uses only a logarithmic number of qubits but more gates to implement oracles. We opt to use the latter.
The binary encoding stores each bosonic mode $\lambda\in[-\Lambda,\Lambda]$ as a binary representation of the integer $\lambda$, using a register of size $\lceil\log(\Lambda)\rceil+1$, where the additional qubit stores the sign of the integer.
Each vertex in the $d$-dimensional lattice has $d$ associated edges, each of which hosts a boson. Thus, for a lattice on $N^d$ sites, we require $dN^d(\lceil\log(\Lambda)\rceil+1)$ qubits. Finally, we impose a consistent local ordering of the edges at a site to establish an ordering of all bosons. The full bosonic register can then be written as
\begin{equation}
    |\lambda\rangle =|\lambda_1^1,\lambda_1^2,\lambda_1^3,\ldots,\lambda_1^d,\lambda_2^1,\lambda_2^2,\ldots,\lambda_N^{d-2},\lambda_N^{d-1},\lambda_N^d\rangle\,,
    \label{eq:boson_basis}
\end{equation}
where the subscripts are site labels and the superscripts are edge labels at the corresponding site.

\subsection{Fermion-to-qubit mapping}
Fermions, unlike bosons, do not suffer from the infinite-dimensional Hilbert space, but possess their own technical challenges to overcome.
The traditional Jordan-Wigner encoding~\cite{jordanwigner} maps fermions onto qubits acted on by strings of Pauli operators within the lattice.
In two or more dimensions, one must impose an ordering of the fermions to generate the strings of Pauli operators, and this leads to non-local operators with weights scaling as $O(N)$. However, in order to invoke the HHKL decomposition~\cite{hhkl2021}, we must encode the fermions in a way which preserves their geometric locality.

\begin{figure}[h]
    \centering
    \subfigure[]{
        \includegraphics[width=0.5\textwidth]
        {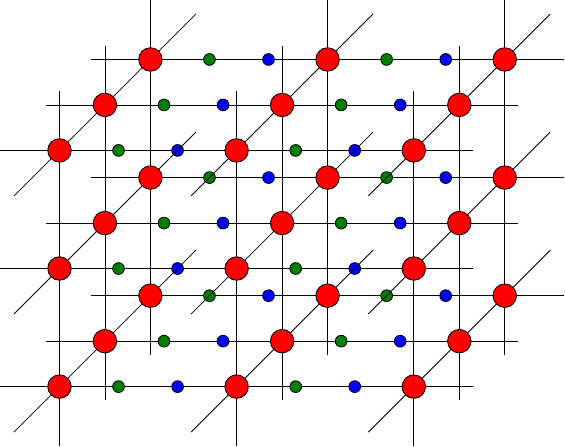}\label{fig:GSE}}
    \hfill
    \subfigure[]{
        \includegraphics[width=0.4\textwidth]
        {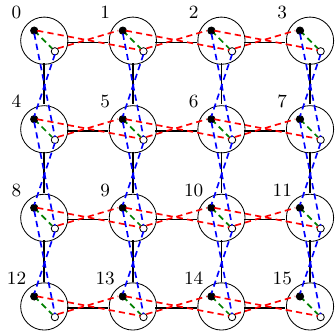}\label{fig:VC_encoding}}
        \caption{\label{fig:fermionic_encodings} The fermion-to-qubit encoding schemes. (a) The generalized superfast encoding~\cite{Setia:2019a} for SU(3) in three spatial dimensions on a $3\times 3\times 3$ cubic lattice. Each lattice vertex of degree $2d$ is assigned $d$ qubits to store a single color DOF (large red circle) and a single edge for every vertex is decorated with $n_c-1$ qubits (small green and blue circles) to store the remaining color DOFs. (b) The Verstraete-Cirac encoding~\cite{cirac2005} for U(1) in two spatial dimensions on a $4\times 4$ square lattice. Each lattice site contains one physical fermion (shaded circle) and one auxiliary fermion (open circle). The sites are enumerated according to our imposed ordering and the vertical and horizontal connectivity between sites is depicted by blue and red dashed lines, respectively, between physical and auxiliary fermions.
        }
\end{figure}

One such way of encoding fermions which preserves geometric locality is the Generalized Superfast (GS) encoding~\cite{Setia:2019a}. Given a $d$-dimensional hypercubic lattice, the GS encoding assigns $d$ qubits to each vertex storing the first fermionic color DOF. Then for each vertex, a single (but consistent) edge is chosen and decorated with $n_c-1$ qubits storing the remaining color DOFs. An example of the GS encoding for SU(3) in three dimensions is shown in Fig.~\ref{fig:GSE}.

In order to map fermionic interactions onto these qubits, an associated operator basis must be chosen. To fix an operator basis, we begin with the fermionic creation and annihilation operators $\psi_j^\dagger$ and $\psi_j^\pd$ for each fermionic mode $j$, or more conveniently, with a pair of Majorana operators $\gamma_{j}=\psi_j^\pd+\psi_j^\dagger$ and $\bar{\gamma}_{j}=-i(\psi_j^\pd-\psi_j^\dagger)$. A generating set for the algebra of even operators is then given by the parity operator $V_v$ for each vertex $v$ and the Majorana bilinear $E_{v,v'}$ for each edge $(v,v')$, defined by
\begin{align}
    V_v&=-i\gamma_{v}\bar{\gamma}_{v}=(-1)^{n_v}\,, \\
    E_{v,v'}&=-i\gamma_{v}\gamma_{v'}\,.
\end{align}
To obtain the qubit counterparts to $V_v$ and $E_{v,v'}$, first define for each vertex $v$ a set of $2d$ independent, mutually anti-commuting Pauli operators $c_{v,1},\ldots,c_{v,2d}$. A local edge ordering at each vertex is imposed such that $c_{v,j}$ corresponds to a Pauli operator originating from the $j$-th edge of vertex $v$. Then the parity and bilinear terms generating the even-parity algebra are
\begin{align}
    V_v&=(-i)^{d}\prod_{j=1}^{2d} c_j\,, \\
    E_{v,v'}&=c_{v,j}c_{v',k}=-E_{v',v}\,,
\end{align}
where $(v,v')$ corresponds to an edge between lattice vertices $v$ and $v'$ while $j$ and $k$ correspond to the imposed edge ordering of each vertex such that the $j$-th edge of vertex $v$ is the $k$-th edge of vertex $v'$. Importantly, this encoding uses only one auxiliary qubit per lattice site and can be easily generalized to higher dimensions and to SU(N) LGTs.

An alternative locality-preserving fermionic encoding in two spatial dimensions is the Verstraete-Cirac (VC) encoding~\cite{cirac2005}. The VC encoding assigns an auxiliary fermion for each physical fermion at a site to facilitate fermionic hopping in two spatial dimensions, as opposed to one dimensional fermionic hopping in the Jordan-Wigner encoding. There is one physical fermion, and thus one auxiliary fermion, for each color DOF in the LGT, meaning $n_c$ auxiliary qubits per lattice site are needed. An example of the VC encoding for U(1) in two dimensions is shown in Fig.~\ref{fig:VC_encoding}.

Again, the fermion interactions must be mapped onto qubit operations obeying the fermionic anti-commutation relations, which amounts to fixing an associated Pauli basis. Such a procedure is discussed in Ref.~\cite{cirac2005}, but the resulting Hamiltonian is not provided. We construct the explicit VC-encoded Hamiltonian here. In order to do so, we need to impose an ordering of the physical and auxiliary fermions in our system, namely the ordering shown in Fig.~\ref{fig:VC_encoding}. We first enumerate the lattice sites using a vector $x=(k,l)$ where $k$ is the column index and $l$ is the row index. Then within each site, the fermions are ordered such that they alternate between physical and auxiliary fermions for each color. The resulting basis is
\begin{equation}
    |f\rangle=|f_1^1,\tilde{f}_1^1,f_1^2,\tilde{f}_1^2,\ldots,f_1^{n_c},\tilde{f}_1^{n_c},f_2^1,\tilde{f}_2^1,\ldots,f_N^{n_c},\tilde{f}_N^{n_c}\rangle
    \label{eq:fermion_basis}\,,
\end{equation}
where $\tilde{f}$ indicates an auxiliary fermion, the subscripts are site labels, and the superscripts are color labels.

Next, the Jordan-Wigner transformation is performed on the composite system of physical and auxiliary fermions in order to obtain the encoded Hamiltonian. In particular, the mass term of Eq.~\eqref{eq:KS_mass} transforms as
\begin{equation}
    H_M\to \sum_{k,l}\sum_{a=1}^{n_c}(-1)^{(k,l)}(1-Z_{k,l}^a)\,,
    \label{eq:VC_mass}
\end{equation}
which acts only on the physical fermionic DOFs. On the other hand, the gauge-matter Hamiltonian of Eq.~\eqref{eq:KS_gauge_matter} splits into two terms. First, we have the usual horizontal hopping term which arises in the Jordan-Wigner encoding, but now acts on both physical and auxiliary fermions,
\begin{align}
    H_{GM}^h=\frac{1}{4}\sum_{k, l} \sum_{a, b = 1}^{{n_c}} \left(X_{k+1, l} - iY_{k+1, l}\right)^a U(k+1, l; k, l)^{ab} \left(X_{k, l} + iY_{k, l}\right)^b S(k+1, l; k, l)^{ab}\,,
    \label{eq:VC_gauge_matter_horiz}
\end{align}
where
\begin{equation}
    S(k+1, l; k, l)^{ab} = \tilde{Z}_{k, l}^b Z_{k, l}^{b+1}\cdots Z_{k, l}^{n_c} \tilde{Z}_{k, l}^{n_c} Z_{k+1, l}^{1} \tilde{Z}_{k+1, l}^1\cdots Z_{k+1, l}^{a-1}\tilde{Z}_{k+1, l}^{a-1}
\end{equation}
is just a string of Pauli $Z$ operators between fermions $f_{(k,l)}^b$ and $f_{(k+1,l)}^a$. In addition to the horizontal hopping term, we obtain a vertical hopping term of the form
\begin{equation}
\begin{aligned}
    H_{GM}^v = \frac{1}{4}\sum_{k, l} \sum_{a, b = 1}^{{n_c}} \left(X_{k, l+1} - iY_{k, l+1}\right)^a &U(k, l+1; k, l)^{ab} \left(X_{k, l} + iY_{k, l}\right)^a \\
    &\times(-1)^{l+1} \begin{dcases}
    \tilde{Y}_{k,l}^a\tilde{X}_{k, l+1}^b,& \text{if $k$ is even, }\\
    \tilde{X}_{k,l}^a\tilde{Y}_{k, l+1}^b,& \text{otherwise.}
    \end{dcases}
    \label{eq:VC_gauge_matter_vert}
\end{aligned}
\end{equation}

The electric field term of Eq.~\eqref{eq:KS_electric} and magnetic field term of Eq.~\eqref{eq:KS_magnetic} act purely on the gauge fields, and are therefore left unchanged under the VC encoding. Lastly, there is an additional auxiliary Hamiltonian term
\begin{equation}
\begin{aligned}
    H_{aux}= &\sum_{\substack{k \text{ odd} \\l \text{ odd} \\}}  \sum_{a=1}^{{n_c}} \tilde{Y}_{k,l}\tilde{Y}_{k+1,l}\tilde{Y}_{k,l+1}\tilde{Y}_{k+1, l+1}S(k + 1, l; k, l)^{aa}S(k+1, l+1; k, l+1)^{aa} \\
    +&\sum_{\substack{ k \text{ even} \\ l  \text{ odd}}}  \sum_{a=1}^{{n_c}} \tilde{X}_{k,l}\tilde{X}_{k+1,l}\tilde{X}_{k,l+1}\tilde{X}_{k+1, l+1}S(k + 1, l; k, l)^{aa}S(k+1, l+1; k, l+1)^{aa}\\
    +&\sum_{\substack{k  \text{ odd} \\l  \text{ even}}}  \sum_{a=1}^{{n_c}} \tilde{Y}_{k,l}\tilde{Y}_{k+1,l}\tilde{Y}_{k,l+1}\tilde{Y}_{k+1, l+1}S(k, l; k-1, l)^{aa}S(k+2, l+1; k+1, l+1)^{aa}\\
    +&\mathrlap{
    \sum_{\substack{k \text{ even} \\ l \text{ even}}}  \sum_{a=1}^{{n_c}} \tilde{X}_{k,l}\tilde{X}_{k+1,l}\tilde{X}_{k,l+1}\tilde{X}_{k+1, l+1}S(k, l; k-1, l)^{aa}S(k+2, l+1; k+1, l+1)^{aa}\,,
    }
\end{aligned}
\end{equation}
which is necessary to fix the ground state of the full Hamiltonian after the addition of the auxiliary fermions. However, as we are only interested in studying the dynamics of LGTs, provided our initial state is an eigenstate of $H_{aux}$, it will only contribute an irrelevant global phase to the final state. In fact, the physical states of the LGT must be eigenstates of $H_{aux}$, so we will always be preparing an initial state which is an eigenstate of $H_{aux}$.

Finally, we note that for U(1) in two spatial dimensions the GS and VC encodings require the same number of qubits per site and have Pauli operators of equivalent weight. We first present explicit circuit constructions for U(1) in two dimensions, and for pedagogical reasons use the explicit VC-encoded Hamiltonian we have just constructed. We defer employing the GS until we discuss generalizations to higher dimensions and arbitrary SU(N) LGTs.

\section{U(1) gauge theory in two dimensions \label{sec:4}}
\subsection{Fast-forwarding $H_E$\label{sec:fast-forward}}

Within the interaction picture, the only term of the Hamiltonian that is simulated without block encoding is the electric field term $H_E$ {as it is diagonal in the electric field basis and can be fast-forwarded}~\cite{Gu_2021}. Applying the HHKL algorithm~\cite{hhkl2021}, the electric field term is decomposed spatially into blocks $\mathcal{B}$ and we are only tasked with determining the circuit complexity of time-evolving $H_E^\mathcal{B}$, the electric field within a single block. 

\begin{figure*}
\begin{center}
    \begin{quantikz}
    \lstick[wires=4]{$\ket{0}^{\otimes p}$} & \gate[5, nwires={3}]{U_k^\pd} & \gate{P(-t)} & \gate[5,nwires={3}]{U_k^\dagger} &\qw \rstick[wires=4]{$\ket{0}^{\otimes p}$}\\
    & & \gate{P(-2t)} & & \qw \\
    & & \vdots & & \\
    & & \gate{P(-2^{p-1}t)} & & \qw\\[0.3cm]
    \lstick{$\ket{k}$} & & \qw & & \qw \rstick{$e^{-itk^2}\ket{k}$}
    \end{quantikz}
\end{center}
\caption{\label{fig:fast-forward circuit} A circuit for fast-forwarding the time evolution of the electric field Hamiltonian $H_E^\mathcal{B}$ in U(1). Given an electric field eigenstate $|k\rangle$, the unitaries $U_k$ prepare the eigenvalue $k^2$ on an ancilla register to $p$ bits of precision, then a series of phase gate rotations $P(\theta)$ are performed to various angles $\theta$, and finally the ancilla register is uncomputed.}
\end{figure*}
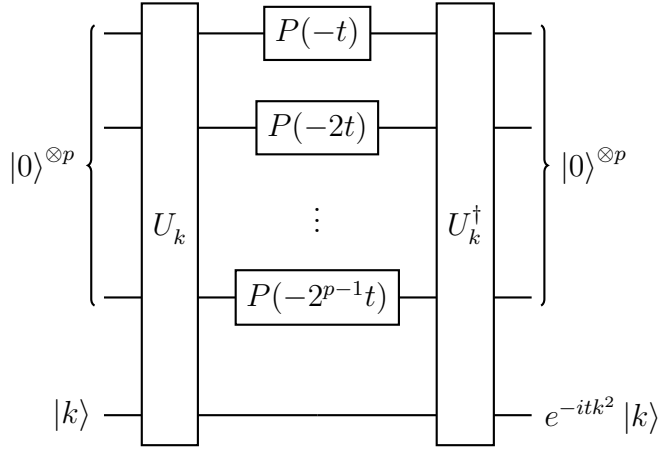

{We efficiently simulate the time evolution from $H_E^\mathcal{B}$ using a general circuit for simulating diagonal operators} in Fig.~\ref{fig:fast-forward circuit}~\cite[Rule 1.6]{childs2004quantum}.
Here, $U_k$ is a unitary which takes an electric field eigenstate and prepares a binary representation of the associated eigenvalue  on an ancilla register to $p$ bits of precision. In general, the eigenvalues of the Casimir operators for LGTs are products of the representation indices, so one way of efficiently implementing $U_k$ is via a Karatsuba-based integer multiplier quantum circuit~\cite{Parent:2017a}. This algorithm requires $4p^2-3p$ Toffoli gates and linear in $p$ ancilla space. However, for the special case of U(1), we note that given an eigenstate $\ket{k}$, we only need to prepare the eigenvalue $k^2$ on the ancilla register to $p=\lceil\log\Lambda^2\rceil=\lceil2\log\Lambda\rceil$ bits of precision. Squaring an integer can be accomplished more efficiently than general multiplication as noted in Ref.~\cite[Lemma 7]{Su:2021a}, requiring only $\lceil\log\Lambda\rceil(\lceil\log\Lambda\rceil-1)$ Toffoli gates. Additionally, $P(\theta)$ is the phase gate which performs rotations about the $z$-axis for each $\theta\in\{-2^{j}t\mod 2\pi\}_{j=0}^{p-1}$. Each phase gate can be implemented to arbitrary precision $\epsilon$ within the Clifford+$\T$ gate set using at most $4\log(1/\epsilon)$ $\T$ gates~\cite{Ross:2016a}. Here, we can set $\epsilon<\min_{0\leq j\leq p-1}\{-2^{j}t\mod 2\pi\}$, meaning the full $\T$-complexity is
\begin{equation}
    \T[e^{-itH_E^\mathcal{B}}]=\Big(8\log\Lambda(\log\Lambda-1/2)+8\log\Lambda\log(1/\epsilon)\Big) \times dN_\mathcal{B}^d\,.
\end{equation}

While the scaling of this circuit holds for arbitrary $\Lambda$ and utilizes the best asymptotically scaling subroutine for implementing $U_k$ to date, the large prefactors make this approach less desirable for practical implementations. In practice $\Lambda$ can often be small, as is the case for the results presented in Ref.~\cite{kan2021} where $\Lambda=10$. For small $\Lambda$, an alternative circuit implementation of $U_k$ can be realized using a quantum lookup table, or equivalently a Quantum Read-Only Memory (QROM).
Here, the possible eigenvalues $k^2$ are classically pre-computed and stored in a database for each $k\in[-\Lambda,\Lambda]$. Upon receiving an index state $\ket{k}$, the QROM loads the associated eigenvalue $k^2$ into the quantum computer. The complexity of implementing a QROM is $4\Lambda-4$~\cite{Babbush:2018a} meaning this alternative approach to fast-forwarding the electric field term scales as
\begin{equation}
    \T[e^{-itH_E^\mathcal{B}}]=\Big(2(4\Lambda-4)+8\log\Lambda\log(1/\epsilon)\Big) \times dN_\mathcal{B}^d\,.
\end{equation}

\begin{figure}
    \centering
    \includegraphics[width=0.75\textwidth]{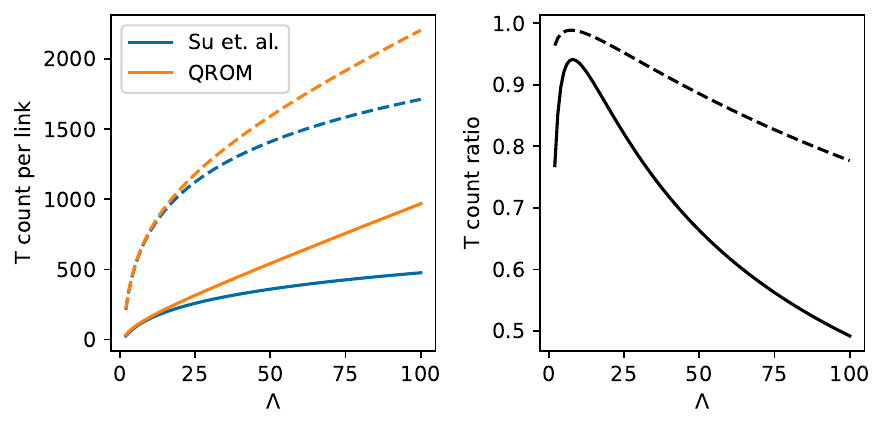}
    \caption{The associated complexity of implementing the unitary $U_k$ for fast-forwarding the electric field term using an arithmetic circuit of Su \textit{et al.} \cite{{Su:2021a}} and QROM {for U(1) gauge theory}.
    Shown are the resources per gauge link, namely $\T[e^{-itH_E^\mathcal{B}}]/(dN_\mathcal{B}^d).$
    Solid lines correspond to $\epsilon = 10^{-1}$ and dashed lines to $\epsilon = 10^{-8},$
    as well as the ratio of the arithmetic $\T$-count to the QROM $\T$-count.
    }
    \label{fig:QROMvsMultiply}
\end{figure}

Fig.~\ref{fig:QROMvsMultiply} shows the complexity of both subroutines for implementing $U_k$ as a function of $\Lambda$.
We see that for both values of $\epsilon$ that the arithmetical implementation is equal or better than using QROM.
The major difference between the two implementations, however, appears at $\Lambda \sim 40$, wherein the arithmetical method is significantly better.
However, in the case of interest $\Lambda = 10$, using either methodology would yield similar estimates for U(1) LGTs.

\subsection{Sparse access oracles}
After fast-forwarding the evolution of $H_E$, the remaining Hamiltonian terms need to be block encoded in order to perform qubitized simulation of the full Hamiltonian. In this section we consider the sparse access oracles, deferring the LCU block encoding oracles to Sec.~\ref{sec:lcu_oracles_u1}.

In general, three oracles are needed to encode a sparse operator:
\begin{align}
    O_r\ket{i}\ket{\ell}&=\ket{i}|c(i,\ell)\rangle\,, \\
    O_c\ket{j}\ket{\ell}&=\ket{j}|r(j,\ell)\rangle\,, \\
    O_H\ket{i}\ket{j}\ket{k}&=\ket{i}\ket{j}|k\oplus H_{ij}\rangle\,,
\end{align}
where $c(i,\ell)$ and $r(j,\ell)$ are functions which return the column and row index of the $\ell$-th non-zero matrix element in row $i$ and column $j$, respectively, and $H_{ij}$ is a binary representation of the Hamiltonian matrix element value. Note that for Hermitian operators, such as Hamiltonians, then we have $O_r=O_c\eqqcolon O_F$ which we will make use of in the remainder of this work.

In order to realize these oracles for the VC-encoded Hamiltonian, we employ some useful features of the Pauli algebra to simplify the structure of the oracles. Generalizing the results of~\cite[Lemma S.1]{Kirby:2021b} to $s$-sparse Hamiltonians, the oracles can be rewritten in the following form suitable for LGT Hamiltonians
\begin{align}
    O_F\ket{f,\lambda}\ket{\ell}&=\ket{f,\lambda}|f\oplus a_\ell,\lambda_\ell^{h,v}-1\rangle\,, \label{eq:sparse_enum}\\
    O_H\ket{f,\lambda}|f',\lambda'\rangle\ket{z}&=\ket{f,\lambda}|f',\lambda'\rangle |z\oplus H_{(f,\lambda),(f',\lambda')}\rangle \label{eq:sparse_value}\,.
\end{align}
Here, the states $\ket{f}$ and $\ket{\lambda}$ store the fermionic and bosonic occupation states as in Eq.~\eqref{eq:fermion_basis} and Eq.~\eqref{eq:boson_basis}, respectively. Further, $\ket{\ell}$ stores a binary representation of an integer indexing the lattice sites and $\ket{z}$ is a register which will store the matrix element value. The matrix element value after the action of a Pauli operator can be expressed as
\begin{equation}
    H_{(f,\lambda),(f',\lambda')}=\begin{cases}
    (-i)^{a_\ell\cdot b_{\ell,j}}(-1)^{b_{\ell,j}\cdot f}, & \textrm{if } |f',\lambda'\rangle=|f\oplus a_\ell,\lambda_\ell^{h,v}-1\rangle\\
    0, & \textrm{otherwise}
    \end{cases},
    \label{eq:matrix_value}
\end{equation}
where $a_\ell$ and $b_{\ell,j}$ are the classically pre-computed length-$2N_\mathcal{B}^d$ symplectic binary representations of a Pauli operator $P=P_1\otimes P_2\otimes \cdots \otimes P_n$, defined by
\begin{equation}
    (a_\ell)_k=\begin{cases}
        0, & \textrm{if }P_k=I,Z \\
        1, & \textrm{if }P_k=X,Y
    \end{cases}\quad\text{ and } \quad
    (b_{\ell,j})_k=\begin{cases}
        0, & \textrm{if }P_k=I,X \\
        1, & \textrm{if }P_k=Y,Z
    \end{cases}.
    \label{eq:symplectic_reps}
\end{equation}
We obtain a vector for each lattice site index $\ell$, and the vectors $b$ are additionally indexed by $j\in[0,3]$ to determine which of the weight-four Pauli operators is applied at the site according to Eq.~\eqref{eq:VC_gauge_matter_horiz} and Eq.~\eqref{eq:VC_gauge_matter_vert}.

We note that the LCU model makes use of an index register $|\ell\rangle$ which iterates over physical sites in a system, while the sparse access model requires $|\ell\rangle$ to index the locations of the non-zero matrix element values. However, here we make use of the fact that the Pauli terms in the Hamiltonian map a fermionic state to at most one other fermionic state, so the sparsity of the Hamiltonian terms is at most $N_\mathcal{B}^d$. The symplectic binary vectors $a_\ell$ are unique for each site, and this allows us to map our lattice site vectors $x=(k,l)$ onto an integer index $\ell$ using the imposed enumeration scheme in Fig.~\ref{fig:VC_encoding}. Therefore, the sparsity index $\ell$ can be more naturally thought of as a lattice site index here.

We have primarily focused on the action of the fermionic operators up until now. This is due to the fact the gauge operators are quite simple in U(1) LGTs, corresponding to incrementing or decrementing the occupied bosonic mode (see Eq.~\eqref{eq:gauge_operator}). Thus, the gauge operators only determine the locations of the non-zero matrix element values, via oracle calls to $O_F$, while oracle calls to $O_H$ will be solely determined by the action of the fermionic operators. We now consider the complexities of constructing these oracles for each of the three Hamiltonian terms.

\subsubsection{Mass Term: $H_M$}
Consider the mass term $H_M$ of Eq.~\eqref{eq:VC_mass} restricted to a geometrically local HHKL block $\mathcal{B}$. For block encoding, the Hamiltonian must be rescaled such that $\|H\|_\textrm{max}\leq 1$, i.e. the maximum matrix element value is at most $|1|$. This amounts to scaling the Hamiltonian by $(N_\mathcal{B}^d/2)^{-1}$. To accomplish this, we will construct oracles for the unnormalized Hamiltonian and at the end of the computation append additional ancilla qubits to the result to bit-shift the encoded values.

\begin{figure*}
\begin{center}
\begin{quantikz}
    \lstick{$\ket{f}$} & \qwbundle{} & \gate[wires=2]{O_F} & \qw & \midstick[2, brackets=none]{=} & \qwbundle{} & \ctrl{1} & \qw\\
    \lstick{$\ket{0}$} & \qwbundle{} & &\qw & & \qwbundle{} & \targ{} & \qw \\
\end{quantikz}
\end{center}
\caption{\label{fig:mass_sparse_enum} The implementation of the enumerator oracle $O_F$ for the mass term $H_M$. Since $H_M$ is diagonal, the row and column indices of the non-zero matrix element values are equivalent.}
\end{figure*}
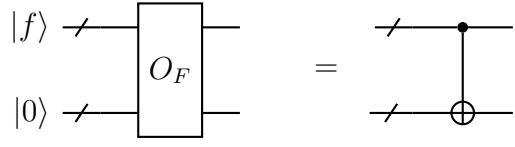

Since the Hamiltonian $H_M$ is diagonal, the enumerator oracle $O_F$ is implemented via a series of CNOT gates which copy the row index to a new register, as shown in Fig.~\ref{fig:mass_sparse_enum}. We omit the bosonic register here for simplicity, as the mass term only acts on the fermionic matter.

\begin{figure*}
\begin{center}
    \begin{quantikz}
        \lstick{$\ket{f}$} & \qwbundle{} & \gate[wires=2]{O_H} & \qw \\
        \lstick{$\ket{0}$} & \qwbundle{} & & \qw
    \end{quantikz}
    \quad = \quad
    \begin{quantikz}[row sep=0.2cm,column sep=0.2cm]
        \lstick[wires=4]{$\ket{f}$} & \ctrl{4} & \qw & \qw & \qw & \qw\\
        & \qw & \ctrl{3} & \qw & \qw & \qw\\
        & & & \vdots &\\
        & \qw & \qw & \qw & \ctrl{1} & \qw \\
        \lstick[wires=4]{$\ket{0}^{\otimes \log (N_\mathcal{B}^d/2)+1}$} & \gate[wires=4,nwires={3}]{+1} & \gate[wires=4,nwires={3}]{-1} & \qw \: \cdots \:\:& \gate[wires=4,nwires={3}]{+1} & \qw \\
        & & & \qw\: \cdots\:\: & & \qw\\
        & & &  \vdots & & \\
        & & & \qw \: \cdots\:\: & & \qw
    \end{quantikz}
\end{center}
\caption{\label{fig:mass_sparse_value} The na\"ive implementation of the matrix element value oracle $O_H$ for the mass term $H_M$. The staggered fermion model imposes that alternating sites occupied by physical fermions amounts to incrementing (+1) or decrementing (-1) the current matrix element value state. The value at the end of the computation is the eigenvalue of the corresponding basis state.
}
\end{figure*}
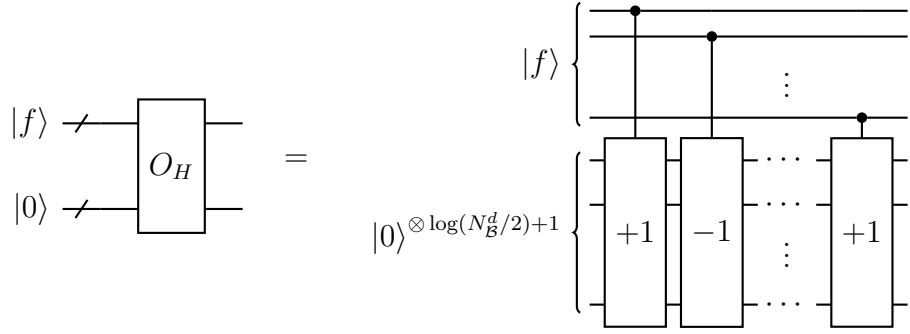

In order to implement the matrix element value oracle, $O_H$, we first note that the spectrum of $H_M$ ranges from $-N_\mathcal{B}^d/2$ to $N_\mathcal{B}^d/2$. Additionally, the staggered fermion structure of the lattice corresponds to incrementing or decrementing the matrix element value on alternating sites, depending on the site occupancy. We store a $p$-bit representation of the matrix element value with the circuit in Fig.~\ref{fig:mass_sparse_value}. Controlled on the occupancy of the \emph{physical} fermion site, the current matrix element value is alternately incremented or decremented, and the final state of the second register stores a binary representation of this value. Finally, this register must be bit-shifted to ensure that $\|H_M\|_\textrm{max}\leq |1|$, but this bit-shift can be implemented by introducing a few additional ancillas.

We further optimize the circuit by defining a Signed-Increment-Decrement (SID) gate.
The SID gate acts on signed integers encoded in a binary representation and will either increment or decrement the value of this integer depending on the state of an ancilla qubit.
This allows us to reduce the number of increments and decrements by a factor of two by first computing the parity of two successive physical fermions in $\ket{f}$. Namely, if the parity is zero, the state remains unchanged, and if the parity is one, either an increment or decrement is performed. Incrementing or decrementing is distinguished by whether the first qubit in a successive pair is one or zero, respectively, so controlling on these registers, the appropriate operation is performed. 

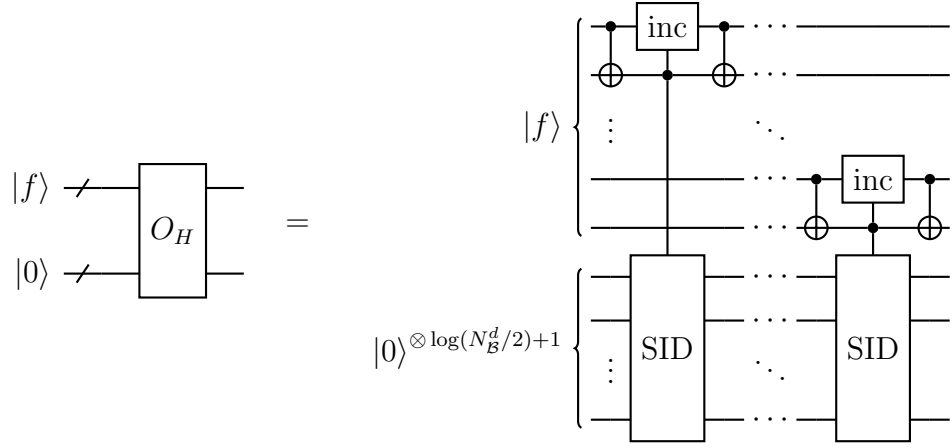
\begin{figure*}
\begin{center}
    \begin{quantikz}
        \lstick{$\ket{f}$} & \qwbundle{} & \gate[wires=2]{O_H} & \qw \\
        \lstick{$\ket{0}$} & \qwbundle{} & & \qw
    \end{quantikz}
    \quad = \quad
    \begin{quantikz}[row sep=0.2cm,column sep=0.1cm]
        \lstick[wires=5]{$\ket{f}$} & \ctrl{1} & \gate{\textrm{inc}} & \ctrl{1} & \qw & \cdots & & \qw &\qw &\qw &\qw \\
        & \targ{} & \ctrl{4}\vqw{-1} & \targ{} & \qw & \cdots & & \qw & \qw & \qw & \qw \\
        & \vdots & & & & \ddots & & & & \\
        & \qw  & \qw &\qw & \qw & \cdots & &\ctrl{1} & \gate{\textrm{inc}} & \ctrl{1} & \qw\\
        & \qw & \qw & \qw & \qw & \cdots & &\targ{} & \ctrl{1}\vqw{-1} & \targ{} & \qw \\
        \lstick[wires=4]{$\ket{0}^{\otimes \log(N_\mathcal{B}^d/2)+1}$} & \qw & \gate[wires=4,nwires={3}]{\text{SID}} & \qw & \qw & \cdots & & \qw & \gate[wires=4,nwires={3}]{\text{SID}} & \qw & \qw\\
        & \qw & &\qw &\qw & \cdots & & \qw & & \qw & \qw\\
        & \vdots & & & & \ddots & & & &\\
        & \qw & & \qw & \qw & \cdots & & \qw & & \qw & \qw\\
    \end{quantikz}
\end{center}
\caption{\label{fig:mass_sparse_value_opt} The optimized implementation of the matrix element value oracle $O_H$ for the mass term $H_M$ using the controlled-SID gate. The parity of successive pairs of physical fermions is computed and temporarily stored on the second qubit of the pair (the $|\text{ctrl}\rangle$ qubit). If this parity is odd, an increment or decrement is performed depending on the state of the first qubit in a pair (the $|\text{inc}\rangle$ qubit). This circuit computes the matrix element value using half of the calls to an increment or decrement circuit using fewer $\T$ gates and only one additional qubit.}
\end{figure*}

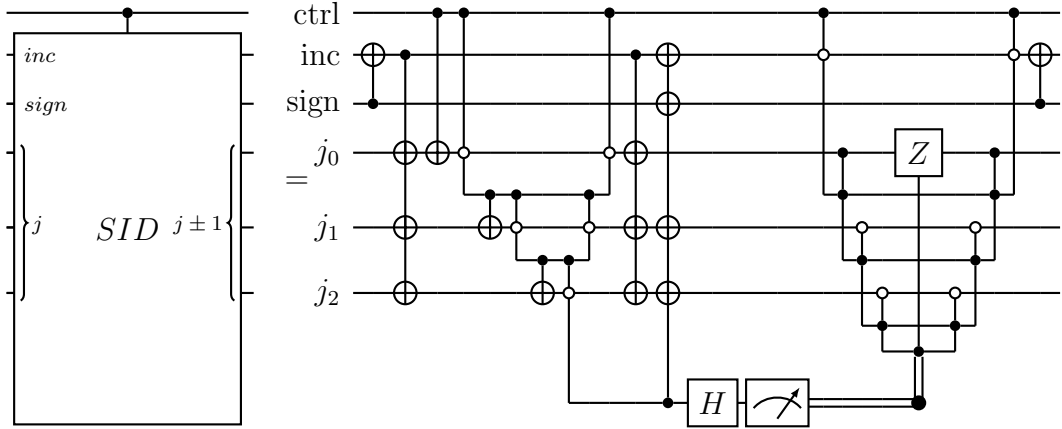
\begin{figure*}
\begin{center}
\begin{quantikz}[row sep=0.2cm,column sep=0.1cm]
& \ctrl{1} & \midstick[10,brackets=none]{=}\qw&
\lstick{ctrl} & \qw & \qw & \ctrl{3} & \ctrl{3} & \qw & \qw & \qw & \qw & \qw & \ctrl{3} & \qw & \qw & \qw & \qw & \ctrl{1} & \qw & \qw & \qw & \qw & \qw & \qw & \qw & \ctrl{1} & \qw & \qw\\
& \gate[wires=10,nwires={4,6,8,9,10}][3cm]{SID}\gateinput{$inc$}  & \qw &
\lstick{inc} & \targ{} & \ctrl{6} & \qw & \qw & \qw & \qw & \qw & \qw & \qw & \qw & \ctrl{6} & \targ{} & \qw & \qw & \octrl{3} & \qw & \qw & \qw & \qw & \qw & \qw & \qw & \octrl{3} & \targ{} & \qw \\
& \qw\gateinput{$sign$} & \qw &
\lstick{sign} & \ctrl{-1} & \qw & \qw & \qw & \qw & \qw & \qw & \qw & \qw & \qw & \qw  & \targ{} & \qw & \qw & \qw & \qw & \qw & \qw & \qw & \qw & \qw & \qw & \qw & \ctrl{-1} & \qw\\
& \qw\gateinput[wires=5]{$j$}\gateoutput[wires=5]{$j\pm1$} & \qw &
\lstick{$j_0$} & \qw & \targ{} & \targ{} & \octrl{1} & \qw & \qw & \qw & \qw & \qw & \octrl{1} & \targ{} & \qw & \qw & \qw & \qw &\ctrl{1} & \qw & \qw & \gate{Z} & \qw & \qw & \ctrl{1} & \qw & \qw & \qw \\
& & &
& & & & & \ctrl{1} & \ctrl{1} & \qw & \qw & \ctrl{1} & \qw & & & & & & \ctrl{2} & \qw & \qw & \qw & \qw & \qw & \ctrl{2} & \qw \\
& \qw & \qw &
\lstick{$j_1$} & \qw & \targ{} & \qw & \qw & \targ{} & \octrl{1} & \qw & \qw & \octrl{1} & \qw & \targ{}  & \targ{} & \qw & \qw & \qw & \qw & \octrl{1} & \qw & \qw & \qw & \octrl{1} & \qw & \qw & \qw & \qw \\
& & &
& & & & & & & \ctrl{1} & \ctrl{1} & \qw & & & & & & & & \ctrl{2} & \qw & \qw & \qw & \ctrl{2} & \qw \\
& \qw & \qw &
\lstick{$j_2$} & \qw & \targ{} & \qw & \qw & \qw & \qw & \targ{} & \octrl{3} & \qw & \qw & \targ{} & \targ{} & \qw & \qw & \qw & \qw & \qw & \octrl{1} & \qw & \octrl{1} & \qw & \qw & \qw & \qw & \qw\\
& & &
& & & & & & & & & & & & & & & & & & \ctrl{1} & \qw & \ctrl{1} & \qw \\
& & &
& & & & & & & & & & & & & & & & & & & \ctrl{-6} & \qw & \\
& & &
& & & & & & & & & \qw & \qw & \qw  & \ctrl{-9} & \gate{H} & \meter{}  & \cw & \cw & \cw & \cw & \cwbend{-1} 
\end{quantikz}
\end{center}
\caption{\label{fig:SID} The controlled Signed-Increment-Decrement (SID) gate.}
\end{figure*}

The qubit overhead can also be reduced by temporarily storing the parity in the second qubit of a successive pair, which then acts as the control register of the SID gate. To avoid using separate subroutines for both the increment and decrement operations, note that the decrement circuit can be obtained from the increment circuit simply by bit-flipping the input state prior to incrementing, so an additional layer consisting of a multi-target CNOT gate is included. An additional qubit must be included to store the sign of the matrix element value. After incorporating all of these techniques, the optimized oracle $O_H$ is shown in Fig.~\ref{fig:mass_sparse_value_opt} and the controlled-SID gate is shown in Fig.~\ref{fig:SID}.

For a register storing the matrix element value of size $\log(N_\mathcal{B}^d/2)$, the controlled-SID gate requires $2\log(N_\mathcal{B}^d/2)+1$ Toffoli gates. Additionally, the controlled-SID gate must be performed $N_\mathcal{B}^d/2$ times. This brings the total $\T$-count of $O_H$ to
\begin{equation}
    \T[O_H]=2N_\mathcal{B}^d(2\log(N_B/2)+1)\,.
    \label{eq:sparse-mass-complexity}
\end{equation}

\subsubsection{Gauge-Matter Term: $H_{GM}$}

For the gauge matter terms, $H_{GM}^h$ and $H_{GM}^v$, both the fermionic and bosonic operators must be considered, though each of these operators can be treated independently. Starting with the fermionic part of $H_{GM}^v$, the symplectic binary vectors $a_\ell$ of the Pauli strings are classically pre-computed according to Eq.~\eqref{eq:symplectic_reps}.

\begin{figure*}
\begin{center}
    \begin{quantikz}
        \lstick{$\ket{f}$} & \qwbundle{} & \gate[wires=2]{O_F} & \qw \\
        \lstick{$\ket{0}$} & \qwbundle{} & & \qw
    \end{quantikz}
    \quad=\quad
    \begin{quantikz}[column sep=0.1cm, row sep=0.25cm]
        \lstick[wires=7]{$\ket{f}$}  & \qw & \qw & \qw & \qw & \qw & \qw & \qw & \ctrl{12} & \qw & \qw & \qw &\qw &\qw &\qw \\
         & \qw & \qw & \qw & \qw & \qw & \qw & \qw & \qw & \ctrl{12} & \qw & \qw & \qw &\qw &\qw \\
         & \qw & \qw & \qw & \qw & \qw & \qw & \qw & \qw & \qw & \ctrl{12} & \qw &\qw & \qw & \qw\\
         & \qw & \qw & \qw & \qw & \qw & \qw & \qw & \qw & \qw & \qw &\ctrl{12} & \qw &  \qw &\qw\\
        \vdots & & &\\
        & \qw & \qw & \qw & \qw & \qw & \qw & \qw &\qw &\qw &\qw &\qw &\ctrl{12} & \qw & \qw\\
         & \qw & \qw & \qw & \qw & \qw & \qw & \qw & \qw &\qw & \qw &\qw &\qw &\ctrl{12} & \qw\\
         \lstick{$\ket{\lambda}$} & \qw & \qw &\qw & \qw & \qw & \qw & \qw & \qw & \qw & \qw &  \qw & \qw & \qw & \qw \\
        \lstick[wires=4]{$\ket{\ell}$}  & \gate[wires=11, nwires={3,9}]{\textrm{QROM}} & \qw & \qw & \qw & \qw & \qw &\qw &\qw &\qw &\qw &\qw &\qw &\qw &\qw\\
        & & \qw & \qw & \qw & \qw & \qw & \qw & \qw &\qw &\qw &\qw &\qw &\qw &\qw\\
         \vdots &\\
         & & \qw & \qw & \qw & \qw & \qw &\qw &\qw &\qw &\qw &\qw &\qw &\qw &\qw\\
        \lstick[wires=7]{$\ket{0}^{\otimes 2N_\mathcal{B}^d}$}  &  & \ctrl{7} & \ctrl{7} & \qw & \qw & \qw & \qw & \targ{} & \qw &\qw &\qw & \qw &\qw &\qw 
        \\
         & & \qw & \qw & \qw & \qw & \qw & \qw & \qw & \targ{} & \qw & \qw & \qw & \qw &\qw \\
         & \qw & \qw & \qw & \ctrl{5} & \ctrl{5} & \qw & \qw & \qw & \qw & \targ{} & \qw & \qw &\qw &\qw\\
         & \qw & \qw & \qw & \qw & \qw & \qw & \qw & \qw & \qw & \qw & \targ{} & \qw & \qw & \qw\\
         \vdots &\\
         & \qw & \qw & \qw & \qw & \qw & \ctrl{2} & \ctrl{2} & \qw & \qw & \qw & \qw & \targ{} & \qw & \qw\\
        & & \qw & \qw & \qw & \qw & \qw & \qw & \qw &\qw &\qw &\qw &\qw & \targ{} & \qw\\
        \lstick{$\ket{0}$} & \qw  & \octrl{2} & \targ{} & \octrl{4} & \targ{} & \octrl{7} & \targ{} & \qw & \qw &\qw & \qw & \qw & \qw & \qw \rstick{$\ket{0}$}\\
         \lstick[wires=7]{$\ket{0}^{\otimes 2N_\mathcal{B}^d}$} & \qwbundle{\log\Lambda} & \qw & \qw & \qw & \qw & \qw & \qw & \qw & \qw & \qw & \qw & \qw & \qw & \qw 
         \\
        & \qwbundle{\log\Lambda}  & \gate{\text{SID}} & \qw & \qw & \qw & \qw & \qw & \qw & \qw & \qw & \qw & \qw & \qw & \qw \\
        & \qwbundle{\log\Lambda}  & \qw & \qw & \qw & \qw & \qw & \qw & \qw & \qw & \qw & \qw & \qw & \qw & \qw \\
        & \qwbundle{\log\Lambda} & \qw & \qw & \gate{\text{SID}} & \qw & \qw & \qw & \qw & \qw & \qw & \qw & \qw & \qw & \qw \\
        \vdots & \\
        & \qwbundle{\log\Lambda}  & \qw & \qw & \qw & \qw & \qw & \qw & \qw & \qw & \qw & \qw & \qw & \qw & \qw \\
        & \qwbundle{\log\Lambda}  & \qw & \qw & \qw & \qw & \gate{\text{SID}} & \qw & \qw & \qw & \qw & \qw & \qw & \qw & \qw \\
    \end{quantikz}
\end{center}
\caption{\label{fig:vert_gm_sparse_enum} The implementation of the enumerator oracle $O_F$ for the vertical gauge-matter Hamiltonian $H_{GM}^v$. First, the QROM loads in the vector $a_\ell$ controlled on the index register $\ket{\ell}$. Then the action of the bosonic operators is performed. Based on the ordering imposed by the VC-encoding, the first non-zero bit in $a_\ell$ indicates the physical site that the link originates from and the third non-zero bit indicates the physical site at which the link terminates (the second and fourth non-zero bits act on auxiliary fermions and do not affect the bosonic links). Note that imposing an ordering of the vertical links, namely that bosonic operators propagate from a starting site downward, eliminates the need for indexing on both the initial and terminal sites, so we need only control on the first non-zero bit of $a_\ell$. Thus we control on every other bit of $a_\ell$ (corresponding to physical sites) of which there will always be two non-zero bits. To avoid performing two decrements, we introduce a flag qubit. If the flag qubit is off, then the decrement is performed on the corresponding bosonic mode $\lambda_\ell^v$ (as will be the case for the first non-zero bit in $a_\ell$) and a successive CNOT flips the state of the flag qubit such that the second non-zero bit of $a_\ell$ does not trigger a decrement. Finally, the fermionic action is performed via a series of CNOT gates between $\ket{f}$ and $\ket{a_\ell}$, as given in Eq.~\eqref{eq:sparse_enum}.}
\end{figure*}
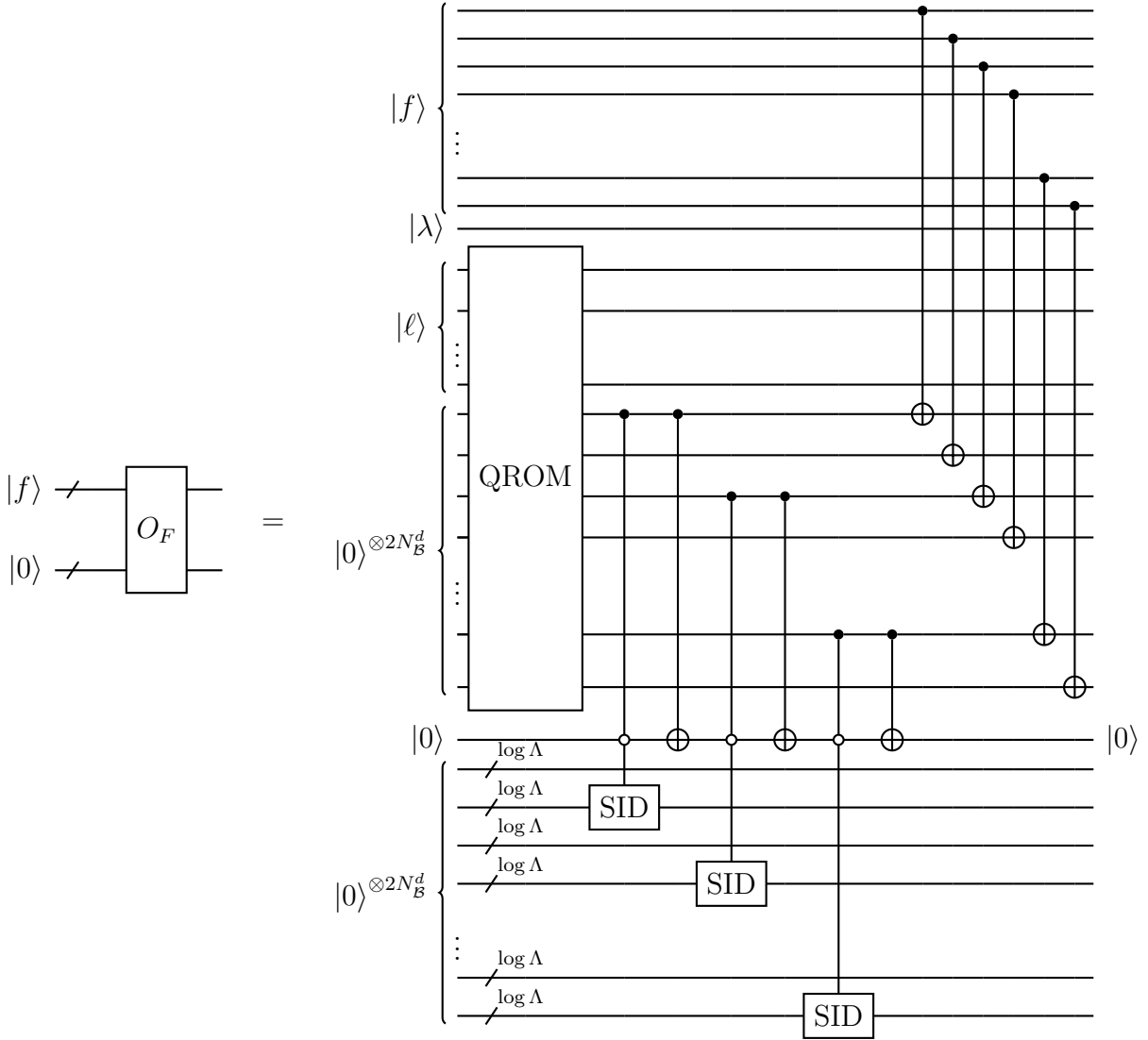

The enumerator oracle $O_F$ is then realized using these symplectic binary vectors which map a fermionic occupancy state onto at most one other fermionic occupancy state by tracking the action of the Pauli operators. The bosonic operators also map a bosonic state to at most one other bosonic state, whose action is simply to increment or decrement a corresponding bosonic mode. Finding the correct final states under the action of the Hamiltonian terms will give us the correct locations of the non-zero matrix elements. The classically pre-computed binary symplectic vectors are loaded in via a QROM~\cite{Babbush:2018a}, just as the eigenvalues were for the electric field term in Sec.~\ref{sec:fast-forward}. Using these vectors, the corresponding actions of the fermionic and bosonic operators of Eq.~\eqref{eq:sparse_enum} can be constructed in a quantum circuit. The explicit circuit is shown in Fig.~\ref{fig:vert_gm_sparse_enum}.

The overall complexity of this circuit will consist of the $\T$-counts from both the QROM and the multi-controlled SID gates. The QROM requires only $4N_\mathcal{B}^d-4$ $\T$ gates~\cite{Babbush:2018a}. To obtain the remaining complexity, we promote each gate in the controlled-SID circuit of Fig.~\ref{fig:SID} by one level of control. Combining these two results, the associated $\T$-count is
\begin{equation}
    \T[O_F]=(4N_\mathcal{B}^d-4)+4N_\mathcal{B}^d(2\log\Lambda+4)=8N_\mathcal{B}^d\log\Lambda+20N_\mathcal{B}^d-4\,.
\end{equation}

\begin{figure*}
\begin{center}
    \begin{quantikz}
        \lstick{$\ket{f}$} & \qwbundle{} & \gate[wires=2]{O_H} & \qw \\
        \lstick{$\ket{0}$} & \qwbundle{} & & \qw
    \end{quantikz}
    \quad = \quad
\begin{quantikz}
    \lstick{$\ket{\ell}$} & \qwbundle{\log N_\mathcal{B}^d} & \qw & \qw & \gate[wires=3]{\textrm{QROM}} & \qw & \qw & \qw & \qw \\
    \lstick{$\ket{+}$} & \qwbundle{2} & \qw & \qw & & \qw & \qw & \qw & \qw \\
    \lstick{$\ket{0}$} & \qwbundle{2N_\mathcal{B}^d} & \qw & \qw &  & \ctrl{1} & \qw & \ctrl{3} & \qw\\
    \lstick{$\ket{f}$} & \qwbundle{2N_\mathcal{B}^d} & \qw & \qw & \qw & \ctrl{4} & \ctrl{2} & \qw & \qw\\
    \lstick{$\ket{\lambda}$} & \qwbundle{2N_\mathcal{B}^d\log\Lambda} & \qw & \qw & \qw & \qw & \qw & \qw &\qw \\
    \lstick{$|f'\rangle=\ket{f\oplus a_\ell}$} & \qwbundle{2N_\mathcal{B}^d} & \qw & \qw & \qw & \qw & \targ{} & \ctrl{3} & \qw\\
    \lstick{$|\lambda'\rangle=|\lambda_\ell^{h,v}-1\rangle$} & \qwbundle{2N_\mathcal{B}^d\log\Lambda} & \qw &\qw & \qw & \qw & \qw & \qw & \qw \\
    \lstick[wires=2]{$\ket{z}$} & \qw & \qw & \qw & \qw  &\targ{} & \qw & \qw & \qw \\
    & \qw & \qw & \qw & \qw & \qw & \qw & \targ{} & \qw
\end{quantikz}
\end{center}
\caption{\label{fig:vert_gm_sparse_value} The implementation of the matrix element value oracle $O_H$ for the vertical gauge-matter Hamiltonian $H_{GM}^v$. First, the same index register $\ket{\ell}$ used in $O_F$ is used to index the QROM which ensures that $a_\ell$ and $b_\ell$ will correspond to the same site $\ell$. An additional two qubits are prepared in the $\ket{+}$ state to index the type of Pauli operator acting on site $\ell$. Then the corresponding binary symplectic representation $b_{\ell,j}$ is loaded in. With this vector, the matrix element value can computed from $H_{f,f\oplus a_\ell}=(-1)^{b_{\ell,j}\cdot f}(-i)^{a_\ell\cdot b_{\ell,j}}$. To do so, the value $b_{\ell,j}\cdot f$ is computed (for each $j$) and stored in the first qubit of $\ket{z}$. In order to compute $a_\ell\cdot b_{\ell,j}$, an XOR is performed between $f$ and $f\oplus a_\ell$, returning $a_\ell$. Then $a_\ell\cdot b_{\ell,j}$ is computed and stored on the second qubit of $\ket{z}$. The state of $\ket{z}$ is then mapped uniquely to the matrix element value $\pm 1$ or $\pm i$.}
\end{figure*}
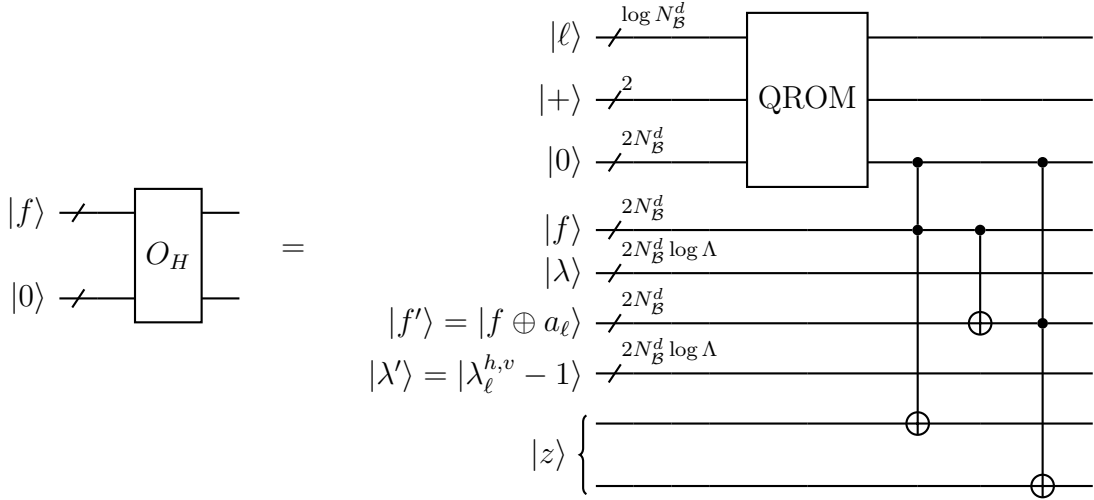

The matrix element value oracle $O_H$ takes a similar form, but now relies on both of the classically pre-computed symplectic binary representations, $a_\ell$ and $b_{\ell,j}$. The vectors $a_\ell$ have already been loaded in as part of the implementation of $O_F$, and we will reuse them here. The vectors $b_{\ell,j}$ are now loaded in via a QROM with two additional qubits in the index register such that a unique vector $b_{\ell,j}$ is loaded for every possible combination of lattice site $\ell$ and weight-four Pauli term $j\in[0,3]$. This amounts to repeating the QROM subroutine four times, one for each unique Pauli term in the Hamiltonian of Eq.~\eqref{eq:VC_gauge_matter_vert}. The matrix element value of Eq.~\eqref{eq:matrix_value} is then determined by computing the bitwise AND operation between two sets of registers. The circuit construction is given in Fig.~\ref{fig:vert_gm_sparse_value}.

The $\T$-complexity results from a call to the QROM as well as two series of Toffoli gates to compute the bitwise AND between the two sets of registers. Combining these two pieces, the $\T$-count is
\begin{equation}
    \T[O_H]=4(4N_\mathcal{B}^d-4)+3+4(4N_\mathcal{B}^d)=32N_\mathcal{B}^d-13\,.
\end{equation}

Finally, we note that the oracle implementations for the horizontal gauge-matter term $H_{GM}^h$ will follow this structure exactly, and can be implemented using equivalent $\T$-counts. Not surprisingly, this is a result of the similar structure in the terms for both the vertical and horizontal gauge-matter Hamiltonians.

\subsubsection{Gauge Term: $H_B$}

For the magnetic field term $H_B$ in U(1), the trace operation of Eq.~\eqref{eq:KS_magnetic} can be ignored as there is only one associated color. Each Hamiltonian term is the product of four bosonic operators forming a plaquette. The implementation of the bosonic operators was discussed in the previous section when considering the gauge-matter interaction, so the circuit for the magnetic field Hamiltonian will just consist of a product of four controlled-SID gates, each acting on their respective bosonic modes.

As mentioned previously, the enumerator oracle $O_F$ is realized by the action of the Hamiltonian terms on the basis states, as this yields the locations of the non-zero matrix elements. Imposing an enumeration scheme such that the lattice site index $\ell$ fixes the upper left vertex of a plaquette uniquely identifies which bosonic modes are acted on by each plaquette term $P_\Box^{(\ell)}$. Given that $\ket{\ell}$ is a superposition over binary encoded states, the na\"ive way to implement $O_F$ is via a series of multi-controlled gates in which there is a unique operation for every possible binary configuration of the controls. The Clifford+$\T$ decomposition for such a circuit is not optimal. 

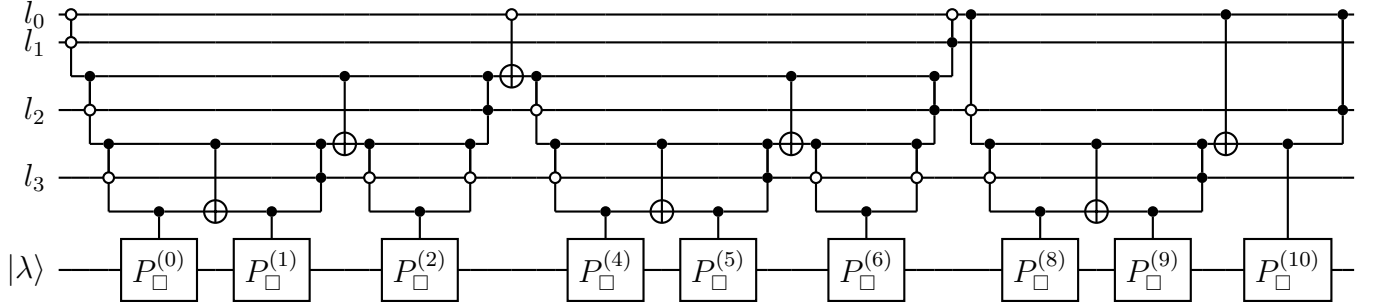
\begin{figure*}
\begin{center}
\begin{quantikz}[column sep=0.08cm, row sep=0.2cm]
\lstick{$l_0$} & \octrl{1} & \qw & \qw & \qw & \qw & \qw & \qw & \qw & \qw & \qw & \qw & \qw & \octrl{2} & \qw & \qw & \qw & \qw & \qw & \qw & \qw & \qw & \qw & \qw & \qw & \octrl{1} & \ctrl{3} & \qw & \qw & \qw & \qw & \qw & \ctrl{4} & \qw & \ctrl{3} & \qw & \\
\lstick{$l_1$} & \octrl{-1} \vqw{1} & \qw & \qw & \qw & \qw & \qw & \qw & \qw & \qw & \qw & \qw & \qw & \qw & \qw & \qw & \qw & \qw & \qw & \qw & \qw & \qw & \qw & \qw & \qw & \ctrl{-1} \vqw{1} & \qw & \qw & \qw & \qw & \qw & \qw & \qw & \qw & \qw & \qw & \\
& & \ctrl{1} & \qw & \qw & \qw & \qw & \qw & \ctrl{2} & \qw & \qw & \qw & \ctrl{1} & \targ{} & \ctrl{1} & \qw & \qw & \qw & \qw & \qw & \ctrl{2} & \qw & \qw & \qw & \ctrl{1} & \qw & & & & & & & & & & & \\
\lstick{$l_2$} & \qw & \octrl{-1} \vqw{1} & \qw & \qw & \qw & \qw & \qw & \qw & \qw & \qw & \qw & \ctrl{-1} \vqw{1} & \qw & \octrl{-1} \vqw{1} & \qw & \qw & \qw & \qw & \qw & \qw & \qw & \qw & \qw & \ctrl{-1} \vqw{1} & \qw & \octrl{-3} \vqw{1} & \qw & \qw & \qw & \qw & \qw & \qw & \qw & \ctrl{-3} \vqw{1} & \qw & \\
& & & \ctrl{1} & \qw & \ctrl{2} & \qw & \ctrl{1} & \targ{} & \ctrl{1} & \qw & \ctrl{1} & \qw & & & \ctrl{1} & \qw & \ctrl{2} & \qw & \ctrl{1} & \targ{} & \ctrl{1} & \qw & \ctrl{1} & \qw & & & \ctrl{1} & \qw & \ctrl{2} & \qw & \ctrl{1} & \targ{} & \ctrl{3} & \qw & & \\
\lstick{$l_3$} & \qw & \qw & \octrl{-1} \vqw{1} & \qw & \qw & \qw & \ctrl{-1} \vqw{1} & \qw & \octrl{-1} \vqw{1} & \qw & \octrl{-1} \vqw{1} & \qw & \qw & \qw & \octrl{-1} \vqw{1} & \qw & \qw & \qw & \ctrl{-1} \vqw{1} & \qw & \octrl{-1} \vqw{1} & \qw & \octrl{-1} \vqw{1} & \qw & \qw & \qw & \octrl{-1} \vqw{1} & \qw & \qw & \qw & \ctrl{-1} \vqw{1} & \qw & \qw & \qw & \qw & \\
& & & & \ctrl{1} & \targ{} & \ctrl{1} & \qw & & & \ctrl{1} & \qw & & & & & \ctrl{1} & \targ{} & \ctrl{1} & \qw & & & \ctrl{1} & \qw & & & & & \ctrl{1} & \targ{} & \ctrl{1} & \qw & & & & & \\
\lstick{\ket{\lambda}} & \qw & \qw & \qw & \gate{P_\Box^{(0)}} & \qw & \gate{P_\Box^{(1)}} & \qw & \qw & \qw & \gate{P_\Box^{(2)}} & \qw & \qw & \qw & \qw & \qw & \gate{P_\Box^{(4)}} & \qw & \gate{P_\Box^{(5)}} & \qw & \qw & \qw & \gate{P_\Box^{(6)}} & \qw & \qw & \qw & \qw & \qw & \gate{P_\Box^{(8)}} & \qw & \gate{P_\Box^{(9)}} & \qw & \qw & \gate{P_\Box^{(10)}} & \qw & \qw & 
\end{quantikz}

\end{center}
\caption{\label{fig:gauge_sparse_enum} The implementation of the enumerator oracle $O_F$ for the magnetic field term $H_B$. The lattice sites $\ell$ are indexed via unary iteration~\cite{Babbush:2018a}, and the corresponding bosonic operators are applied around the associated plaquette. An example is shown here using the enumeration scheme of the $4\times 4$ lattice in Fig.~\ref{fig:VC_encoding}.
The plaquettes are enumerated by the index of their upper left lattice sites.
}
\end{figure*}

Instead, we make use of the unary iteration subroutine which converts a binary encoding into a successive preparation, then unpreparation, of unary encoded states which only require operations with a single control~\cite{Babbush:2018a}. 
Using unary iteration, each lattice site index is successively controlled on and the associated plaquette operator $P_\Box^{(\ell)}$ is applied. This is accomplished via the circuit in Fig.~\ref{fig:gauge_sparse_enum}.

The complexity of this oracle combines the cost of unary iteration and the cost of implementing four bosonic operators at each lattice site. The full complexity is
\begin{equation}
    \T[O_F]=4N_\mathcal{B}^d+4(4N_\mathcal{B}^d(2\log\Lambda+1))=32N_\mathcal{B}^d\log\Lambda+20N_\mathcal{B}^d\,.
\end{equation}
The value oracle $O_H$ for the magnetic field term is trivial in U(1) as all matrix element values are 1.

\subsubsection{Resource requirements \label{sec:sparse-resources}}

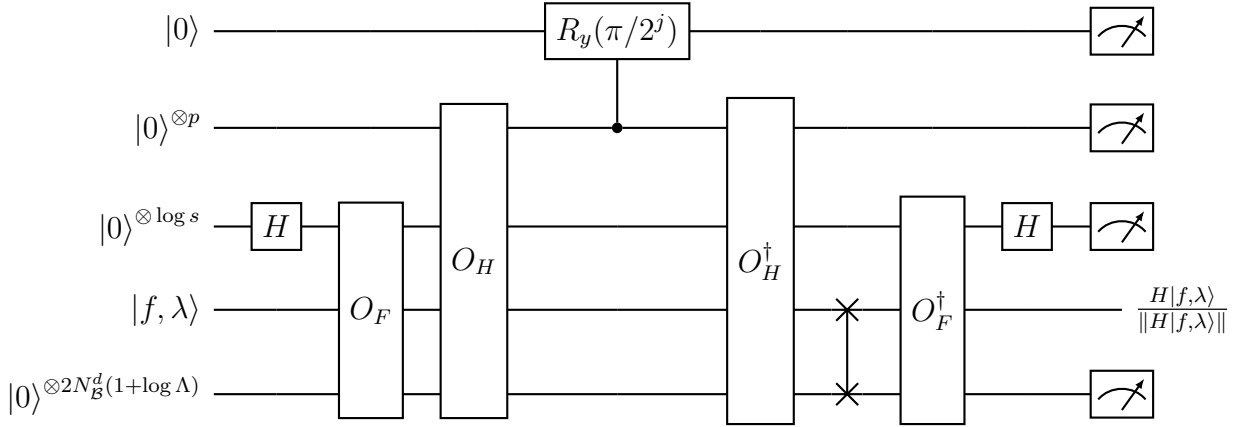
\begin{figure*}
    \centering
    \begin{quantikz}
        \lstick{$\ket{0}$} & \qw & \qw & \qw & \gate{R_y(\pi/2^j)} & \qw & \qw & \qw & \qw & \meter{} \\
        \lstick{$\ket{0}^{\otimes p}$} & \qw & \qw & \gate[wires=4]{O_H} & \ctrl{-1} & \gate[wires=4]{O_H^\dagger} & \qw & \qw & \qw & \meter{} \\
        \lstick{$\ket{0}^{\otimes \log s}$} & \gate{H} & \gate[wires=3]{O_F} & & \qw & & \qw & \gate[wires=3]{O_F^\dagger} & \gate{H} &\meter{}\\
        \lstick{$\ket{f,\lambda}$} & \qw & & & \qw & & \swap{1} & & \qw & \qw \rstick{$\frac{H\ket{f,\lambda}}{\|H\ket{f,\lambda}\|}$}\\
        \lstick{$\ket{0}^{\otimes 2N_\mathcal{B}^d(1+\log\Lambda)}$} & \qw & & & \qw & & \targX{} & & \qw & \meter{}
    \end{quantikz}
    \caption{\label{fig:sparse_block_encoding} A circuit to block encode the Hamiltonian via sparse access oracles~\cite{Lin:2022a}. The top qubit is the signal qubit, the second register stores the matrix element value to $p$ bits of precision. The third register stores a lattice site index, the fourth register stores the Hamiltonian matrix row indices, and the fifth register stores the matrix element column indices. Importantly, this circuit is successful only upon all measurement outcomes being 0.}
\end{figure*}

Given the explicit sparse access oracle constructions, the Hamiltonian interaction terms can be block encoded. This circuit complexity is given in Ref. \cite[Lemma~48]{Gily_n_2019} and a circuit representation is given in Ref.~\cite{Lin:2022a}. In terms of our registers, the block encoding circuit is shown in Fig.~\ref{fig:sparse_block_encoding}.

Since block encoding the Hamiltonian interaction terms is realized via two calls to $O_F$ and $O_H$, as well as a series of controlled rotations, its full complexity is
\begin{equation}
    \T[H]=2\T[O_F]+2 \T[O_H]+ 4p(4\log(1/\epsilon))\, ,
\end{equation}
where $p$ is the number of qubits needed to store the matrix element value and $\epsilon$ is the rotation synthesis error to which a rotation can be approximated in the Clifford+$\T$ gate set, using the optimal bound of Ref.~\cite{Ross:2016a}. For U(1) gauge theory $p= \log_2(N_\mathcal{B}^d/2)$ for storing the terms in $H_M$. Then for each Hamiltonian term, their complexity scales as
\begin{align}
    &\hphantom{\T[H_{GM}^\mathcal{B}]}
    \mathllap{
    \T[H_M^\mathcal{B}]}
    = 4N_\mathcal{B}^d(2\log(N_\mathcal{B}^d/2)+1)+16\log(N_\mathcal{B}^d/2)\log(1/\epsilon) \simeq 8N_\mathcal{B}^d\log(N_\mathcal{B}^d/2)\,, \\
    &\begin{alignedat}{9}
    \T[H_{GM}^\mathcal{B}]&=2(4N_\mathcal{B}^d-4+4N_\mathcal{B}^d(2\log\Lambda+4))+2(4(4N_\mathcal{B}^d-4)+3+4(4N_\mathcal{B}^d)) + 32\log(1/\epsilon)\,, \\
    &\simeq 16N_\mathcal{B}^d\log\Lambda+104N_\mathcal{B}^d \,,
    \end{alignedat}
    \\
    &\begin{alignedat}{9}
    \hphantom{\T[H_{GM}^\mathcal{B}]}
    \mathllap{\T[H_B^\mathcal{B}]}
    &=2(4N_\mathcal{B}^d - 8 + 8(4(N_\mathcal{B}^d)(2\log\Lambda+1))+2N_\mathcal{B}(2\log\Lambda +1) + 16\log(1/\epsilon)
    \\
    &\simeq 64N_\mathcal{B}^d\log\Lambda+72N_\mathcal{B}^d \,.
    \end{alignedat}
\end{align}

The reason that two calls to $O_H$ are made, along with a series of controlled rotations, is to implement an alternative version of the matrix element value oracle
\begin{equation}
    \tilde{O}_H\ket{0}\ket{i}\ket{j}=\left(H_{ij}\ket{0}+\sqrt{1-|H_{ij}|^2}\ket{1}\right)\ket{i}\ket{j}\,,
    \label{eq:alternative-sparse-value}
\end{equation}
which encodes the matrix element value in the amplitude of the signal qubit. While for U(1) it is possible to remove this overhead by implementing Eq.~\eqref{eq:alternative-sparse-value} directly, in general this approach will not be as straightforward. When matrix values are more complicated functions of the system parameters, such as those in SU(2) and SU(3) LGTs, it is more convenient to design an oracle that stores the $p$-bit representation of the matrix element value, as we have done, and then convert to the signal qubit encoding when block encoding via the circuit in Fig.~\ref{fig:sparse_block_encoding}.

\subsection{LCU oracles \label{sec:lcu_oracles_u1}}
Consider the LCU oracles for a Hamiltonian $H=\sum_{\ell=0}^{L-1}\alpha_\ell U_\ell$ with norm $\|H\|\leq \|\vec{\alpha}\|_1=\sum_{\ell=0}^{L-1}|\alpha_\ell|$, defined in Ref.~\cite{Low:2019a} as
\begin{align}
    \textsc{prepare}\ket{0}&=\sum_{\ell=0}^{L-1}\sqrt{\frac{\alpha_\ell}{\|\vec{\alpha}\|_1}}\ket{\ell}, \\
    \textsc{select}\ket{\ell}\ket{\psi}&=\ket{\ell}U_\ell\ket{\psi}.
\end{align}
Since the phases can be absorbed into the unitary operators $U_\ell$, we can assume all coefficients $\alpha_\ell$ are positive and real. Thus, $\textsc{prepare}$ generates a superposition of index states weighted by the Hamiltonian coefficients $\alpha_\ell$ and $\textsc{select}$ applies the unitary term $U_\ell$ controlled on the index state $\ket{\ell}$.

\subsubsection{Mass Term: $H_M$}
For the mass term $H_M$ in U(1), first rewrite the Hamiltonian of Eq.~\eqref{eq:VC_mass} in a form more suitable to the LCU oracles,
\begin{equation}
    H_M=\frac{1}{2}\sum_{\ell=0}^{N_\mathcal{B}^d-1} (-1)^{\ell+1}Z(\ell)\,,
\end{equation}
where the lattice vector indices $x=(k,l)$ have been mapped onto scalar indices $\ell$ with the ordering imposed by the VC-encoding, much like the indexing scheme for the sparse access oracles. The phases $(-1)^{\ell+1}$ can all be absorbed into the unitaries such that $\alpha_\ell=1/2$ for all $\ell$. Then the action of the \textsc{prepare} oracle is
\begin{equation}
    \textsc{prepare}\ket{0}=\frac{1}{\sqrt{N_\mathcal{B}^d}}\sum_{\ell=0}^{N_\mathcal{B}^d-1}\ket{\ell},
\end{equation}
corresponding to a uniform superposition over the site indices. This circuit is implemented with a single layer of Hadamard gates, and thus has a trivial complexity. As we will see, \textsc{prepare} will be trivial for each Hamiltonian interaction term in U(1), making it a desirable encoding model for this problem.

\begin{figure*}
    \includegraphics[width=\textwidth]{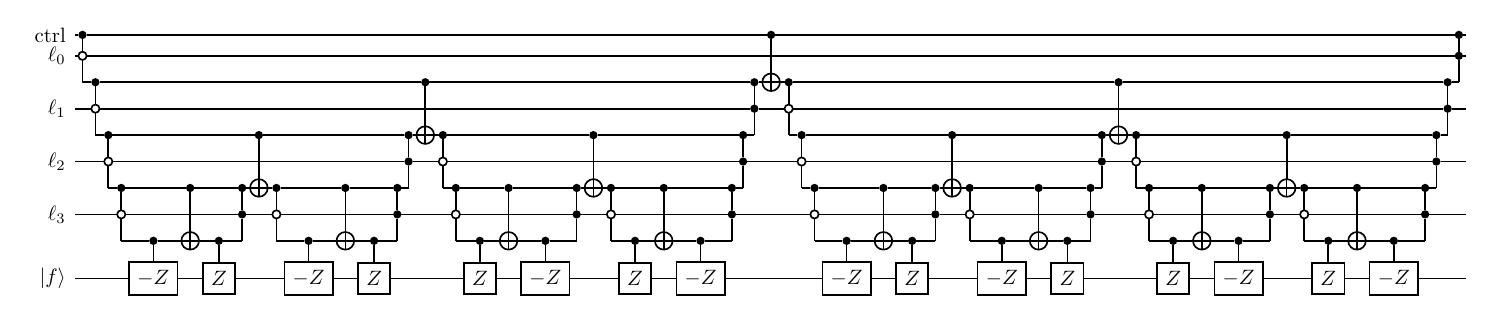}
    \caption{\label{fig:mass_LCU_select} The implementation of \textsc{select} for the mass term $H_M$ using unary iteration~\cite{Babbush:2018a}. Controlled on the iteratively produced unary encoded site indices $|\ell\rangle$, the controlled $\pm Z$ gates are performed on the corresponding physical fermion sites.}
\end{figure*}

On the other hand, the \textsc{select} oracle is implemented via operations controlled on the index states $\ket{\ell}$. As mentioned previously, rather than performing a series of multi-controlled gates for all possible binary configurations of the index state, we use the unary iteration circuit of Ref.~\cite{Babbush:2018a}. Then the action of \textsc{select} is to apply $\pm Z$ to the physical fermion sites controlled on the unary encoded index states. This is accomplished via the circuit in Fig.~\ref{fig:mass_LCU_select}.
The total $\T$-gate count of \textsc{SELECT} circuit comes from the unary iteration subroutine, so
\begin{equation}
    \T[\textsc{select}]=4N_\mathcal{B}^d-4\,,
\end{equation}
and we obtain an $O(\log N_\mathcal{B}^d)$ improvement over the sparse access encoding of the mass term in Eq.~\eqref{eq:sparse-mass-complexity}.

\subsubsection{Gauge-Matter Term: $H_{GM}$}

The vertical gauge-matter term $H_{GM}^v$ of Eq.~\eqref{eq:VC_gauge_matter_vert} is also naturally suitable to the \textsc{prepare} and \textsc{select} oracles. Again imposing an enumeration of the lattice sites fixes the locations of the fermions which can be acted on by the gauge-matter term. This then fixes the structure of the associated Pauli operators and the overall phase of the Pauli operator, i.e. the row and column parities are uniquely determined by $\ell$ under a specified enumeration scheme. Then the action of the \textsc{select} oracle is

\begin{equation}
    \textsc{select}\ket{A,B,\ell}\ket{\psi}=\ket{A,B,\ell}\otimes\begin{cases}
        \pm X_\ell \tilde{Y}_\ell X_{\ell'} \tilde{X}_{\ell'}U(\ell';\ell)\ket{\psi}, & A \land B\land \ell\textrm{ is even} \\
        \pm i Y_\ell \tilde{Y}_\ell X_{\ell'} \tilde{X}_{\ell'} U(\ell';\ell)\ket{\psi}, &  A \land \neg B\land \ell\textrm{ is even} \\
        \mp i X_\ell \tilde{Y}_\ell Y_{\ell'} \tilde{X}_{\ell'} U(\ell';\ell) \ket{\psi}, & \neg A\land B\land \ell\textrm{ is even} \\
        \pm Y_\ell \tilde{Y}_{\ell} Y_{\ell'} \tilde{X}_{\ell'} U(\ell';\ell)\ket{\psi}, & \neg A \land \neg B \land \ell\textrm{ is even} \\
        \pm X_\ell \tilde{X}_\ell X_{\ell'} \tilde{Y}_{\ell'}U(\ell';\ell)\ket{\psi}, & A \land B\land \ell\textrm{ is odd} \\
        \pm i Y_\ell \tilde{X}_\ell X_{\ell'} \tilde{Y}_{\ell'} U(\ell';\ell)\ket{\psi}, &  A \land \neg B\land \ell\textrm{ is odd} \\
        \mp i X_\ell \tilde{X}_\ell Y_{\ell'} \tilde{Y}_{\ell'} U(\ell';\ell) \ket{\psi}, & \neg A\land B\land \ell\textrm{ is odd} \\
        \pm Y_\ell \tilde{X}_{\ell} Y_{\ell'} \tilde{Y}_{\ell'} U(\ell';\ell)\ket{\psi}, & \neg A \land \neg B \land \ell\textrm{ is odd} \\
    \end{cases}
    \label{eq:LCU_select_gm}
\end{equation}
 where $\ell'$ is a function of $\ell$ determined \emph{a priori} from the lattice enumeration and the phase of each term also has a known dependence on $\ell$. Here, $A$ and $B$ are single qubit index registers whose states determine which type of Pauli operators are applied. Much like the mass term, since the phases are absorbed into the Pauli terms, then $\alpha_\ell=1/4$ for all $\ell$. The \textsc{prepare} oracle takes the form
\begin{equation}
    \textsc{prepare}\ket{0}=\frac{1}{2\sqrt{N_B}}\sum_{\ell=0}^{N_\mathcal{B}^d-1} \ket{+}_A\ket{+}_B\ket{\ell}\,,
\end{equation}
which is again just the uniform superposition state and can be implemented trivially.

\begin{figure*}
\includegraphics[width=\textwidth]{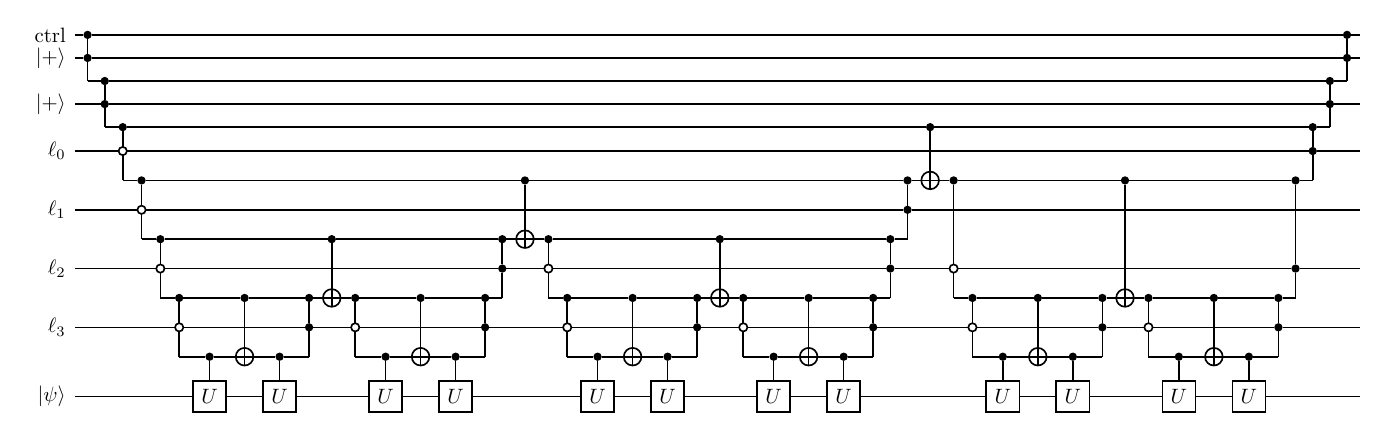}
\caption{\label{fig:LCU_vertical_unary} The partial implementation of \textsc{select} for the gauge-matter term $H_{GM}$. This example shows the first configuration for $\ket{A,B}=\ket{11}$ and thus will apply $U=\pm X_\ell \tilde{Y}_\ell X_{\ell'} \tilde{X}_{\ell'}U(\ell';\ell)$ or $U=\pm X_\ell \tilde{X}_\ell X_{\ell'}\tilde{Y}_{\ell'}U(\ell',\ell)$ according to Eq.~\eqref{eq:LCU_select_gm} depending on whether $\ell$ is an even or odd site. This circuit will then be repeated for the remaining three configurations of $\ket{A,B}$ to apply the rest of the terms.}
\end{figure*}

To implement \textsc{select}, unary iteration is again used to successively generate the index states $\ket{\ell}$ and the additional index qubits $A$ and $B$ will not be iterated over in unary iteration, but instead index the Pauli operator applied at each site. Thus, unary iteration must be repeated over the lattice sites four times, once for each configuration of $\ket{A,B}$. The first iteration of the \textsc{select} oracle is shown in Fig.~\ref{fig:LCU_vertical_unary}.

The full complexity is once again determined from both unary iteration as well as the cost of implementing the controlled unitary operations at each lattice site. However, while the fermionic unitaries can be implemented trivially, to implement the gauge operator we need to implement a controlled-SID gate once again. The complexity is thus 
\begin{equation}
    \T[\textsc{select}]=4(4N_\mathcal{B}^d-4)+3+4N_\mathcal{B}^d(2\log\Lambda+1)=8N_\mathcal{B}^d\log\Lambda+20N_\mathcal{B}^d-13\,.
\end{equation}

As for the horizontal hopping term $\hat{H}^h_{GM}$, as given in Eq.~\eqref{eq:VC_gauge_matter_horiz}, it takes the same form as the vertical hopping term in the LCU formalism.
The only difference is that the Pauli strings in the \textsc{select} operation can be higher weight for SU(2) and SU(3), but are lower weight for U(1). In either case these do not contribute any additional $\T$ gates.
The total $\T$-gate count for the combined vertical and horizontal hoppings are double that of the horizontal contribution, and we require one additional control qubit $D$ which determines the hopping direction, vertical or horizontal.

\subsubsection{Gauge Term: $H_B$}
The magnetic field term $H_B$ for U(1) amounts to decrementing, or incrementing, the bosonic modes on the links forming a plaquette in the lattice, as was the case for the sparse access oracles. Once again, the lattice structure plays a crucial role in the construction of these oracles as a site index $\ell$ will allow us to uniquely determine the plaquette, and thus the bosonic modes that need to be modified. In particular,
\begin{equation}
    \textsc{select}\ket{\ell}\ket{\psi}=U^h(\ell+1;\ell)U^v(\ell+N_\mathcal{B}^d+1;\ell+1)U^{h\dagger}(\ell+N_\mathcal{B}^d;\ell+N_\mathcal{B}^d+1)U^{v\dagger}(\ell;\ell+N_\mathcal{B}^d)\sket{\psi}\,,
\end{equation}
where $U^{h}(f,i)$ ($U^v(f,i)$) acts on the horizontal (vertical) links oriented from site $i$ to site $f$. The circuit for implementing \textsc{select} will actually be identical to the circuit implementing the sparse access oracle $O_F$ for the magnetic field term in U(1), though an additional control qubit is needed for the LCU implementation.

The \textsc{prepare} oracle does not require $\T$ gates and only involves Hadamards, as we only need to prepare the uniform superposition over the index register $\ket{\ell}$,
\begin{equation}
    \textsc{prepare}\ket{0}=\frac{1}{\sqrt{N_\mathcal{B}^d}}\sum_{\ell=0}^{N_\mathcal{B}^d-1}\ket{\ell}.
\end{equation}

\subsubsection{Resource requirements \label{sec:lcu-resources}}
Block encoding the Hamiltonian using LCU oracles is accomplished via the circuit shown in Fig.~\ref{fig:lcu_block_encoding}, with a $\T$-count of
\begin{equation}
    \T[H]=2\T[\textsc{prepare}]+\T[\textsc{select}]\, , 
\end{equation}
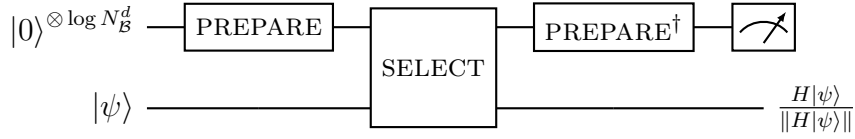
\begin{figure*}
    \centering
    \begin{quantikz}
        \lstick{$\ket{0}^{\otimes\log N_\mathcal{B}^d}$} & \gate{\textsc{prepare}} & \gate[wires=2]{\textsc{select}} & \gate{\textsc{prepare}^\dagger} & \meter{} \\
        \lstick{$\ket{\psi}$} & \qw & & \qw & \qw\rstick{$\frac{H\ket{\psi}}{\|H\ket{\psi}\|}$}
    \end{quantikz}
    \caption{\label{fig:lcu_block_encoding}Block encoding circuit in the LCU model~\cite{childs2012hamiltonian,Lin:2022a}.}
\end{figure*}
such that
\begin{align}
    \T[H_M^\mathcal{B}]&= 4N_\mathcal{B}^d-4\simeq 4N_\mathcal{B}^d\,,\\
    \T[H_{GM}^\mathcal{B}] &= 2(4(4N_\mathcal{B}^d-4)+3+N_\mathcal{B}^d(4(2\log\Lambda+1)))\simeq 16N_\mathcal{B}^d\log\Lambda+40N_\mathcal{B}^d\,, \\
    \T[H_B^\mathcal{B}]&= 4N_\mathcal{B}^d-4+8N_\mathcal{B}^d(4(2\log\Lambda+1))+2N_\mathcal{B}(2\log\Lambda+1)\simeq 32N_\mathcal{B}^d\log\Lambda+36N_\mathcal{B}^d\,.
    \end{align}

\subsection{Resource comparison}
Here we compute the total resource requirements for a two dimensional U(1) LGT Hamiltonian simulation using the interaction picture simulation method, HHKL block decomposition, and the costs for electric field circuits and block encoding circuits determined in Sec.~\ref{sec:sparse-resources} and Sec.~\ref{sec:lcu-resources}.
We conclude this section by comparing to the cost of Trotterized Hamiltonian simulation developed by Kan and Nam~\cite{kan2021}.

We first establish the resource requirements for the total block encoding of $H^{\mathcal{B}}_{M + GM + B}$.
To prepare a linear superposition over our three block encodings $H_M^\mathcal{B}$, $H_{GM}^\mathcal{B}$, and $H_B^\mathcal{B}$ with constants $g_M$, $g_{GM}$, and $g_B$, we use two ancilla registers containing a single-qubit rotation on each ancilla, and controlled application of each of the three block-encoded operators therein.
The cost for doubly-controlled applications of the sparse block encodings are:
\begin{align}
&
\hphantom{\T[CC-H_{GM}^\mathcal{B}]}
\mathllap{\T[CC-H_M^\mathcal{B}]}
= 16N_\mathcal{B}^d+ 4N_\mathcal{B}^d(2\log(N_\mathcal{B}^d/2)+5)+16\log(N_\mathcal{B}^d/2)\log(1/\epsilon) \,,\\
&
\begin{alignedat}{9}
\T[CC-H_{GM}^\mathcal{B}]=40N_\mathcal{B}^d + &2(4N_\mathcal{B}^d+4+4N_\mathcal{B}^d(2\log\Lambda+6))
\\+&2(4(4N_\mathcal{B}^d-4)+5+32N_\mathcal{B}^d) + 32\log(1/\epsilon)\,,     
\end{alignedat}
\\
&
\hphantom{\T[CC-H_{GM}^\mathcal{B}]}
\mathllap{\T[CC-H_B^\mathcal{B}]}=2(4N_\mathcal{B}^d - 4 + 32N_\mathcal{B}^d(2\log\Lambda+1)+2N_\mathcal{B}(2\log\Lambda +1)) + 16\log(1/\epsilon)\,, \\
&\hphantom{\T[CC-H_{GM}^\mathcal{B}]}
\mathllap{
Q}=6N_\mathcal{B}^d+4N_\mathcal{B}^d\log\Lambda+2\log N_\mathcal{B}^d +2\,,
\end{align}
where $Q$ is the number of qubits required, and similarly for the LCU encodings:
\begin{align}
\T[CC-H_M^\mathcal{B}] &=  4N_\mathcal{B}^d + 4\,,\\
\T[CC-H_{GM}^\mathcal{B}] &= 2(4(4N_\mathcal{B}^d + 4) + 3 + 4N_\mathcal{B}^d(2\log\Lambda + 1))\,,\\
\T[CC-H_B^\mathcal{B}] &= 4N_\mathcal{B}^d+4+32N_\mathcal{B}^d(2\log\Lambda+1)+2N_\mathcal{B}(2\log\Lambda+1)\,, \\
Q&=N_\mathcal{B}^d+2N_\mathcal{B}^d\log\Lambda+2\log N_\mathcal{B}^d+2\,.
\end{align}

The total cost of block encoding is then
\begin{equation}
\T[H^\mathcal{B}_{M+GM+B}]\equiv \T[CC-H_M^\mathcal{B}] + \T[CC-H_{GM}^\mathcal{B}] + \T[CC-H_B^\mathcal{B}] + 6\log(1/\epsilon)\,.    
\end{equation}

We note that the addition of the two extra control qubits yields just a constant overhead for LCU due to the presence of unary logic, whereas significant overheads are present for the sparse access model.
This is because the circuits for the sparse access models involve  CNOT gates for logic, which get elevated to multi-controlled Toffoli gates once two additional control lines are added.

\begin{figure}[ht]
        \centering
    \includegraphics[width=\textwidth]{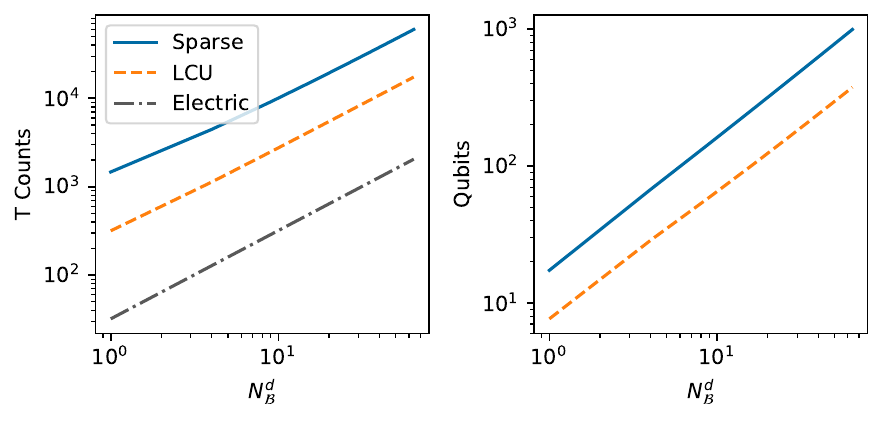}
    \caption{\label{fig:total_counts} The total (a) $\T$-counts and (b) qubit counts needed to block encode the Hamiltonian. The LCU model achieves $\sim 3$x improvement over the sparse model in the $\T$-count and $\sim 3$x improvement in qubit count. Here we present results for $\Lambda$ = 5 and $\epsilon = 10^{-3}$.}
\end{figure}

We present these resource estimates in Fig.~\ref{fig:total_counts}, alongside the cost of implementing the fast-forwarded term $e^{i\tau H_{E}^\mathcal{B}}$.
We find that the LCU block encoding beats the sparse encoding by a factor of $\sim 3$ in $\T$-counts.
Further, the cost of simulating the electric term is an order of magnitude smaller than either block encoding.
This would indicate that the LCU encoding significantly outpaces the sparse encoding in practical applications.
However, as mentioned in Section~\ref{sec:2} on qubitized simulation, the oracle that is queried in the interaction picture simulation is HAM-T, which requires multiple queries to the electric field term but only a single query to the block encoding.
We will see shortly that the overhead of multiple queries eliminates any $\T$-gate reduction from LCU.
We additionally find a factor of $2 - 3$ improvements in qubit counts in the LCU block encoding relative to the sparse encoding.

In implementing HAM-T, we note by Ref.~\cite[Theorem 7]{Low:2019b} it is computed using $\lceil \log_2 \mathcal{M} \rceil$ queries to controlled applications of $e^{iH_E^\mathcal{B} /(2\alpha)}$ where $\alpha = \text{max}_{s \in \{0, T\}}||H_{I}^\mathcal{B}(s)||$ defined in Eq.~\eqref{eq:interaction-block}, and a single application of the block encoding $\T[H_{M+GM+B}^\mathcal{B}].$
Here $\mathcal{M} = 16(\alpha + \alpha_E)/\epsilon$ where $\alpha_E = g_E ||H_E^\mathcal{B}||.$
Noting the coefficients in the Hamiltonian, we can express the norms as $\alpha = 2(g_M + g_{GM} + |g_B|)N_\mathcal{B}^d$, $\alpha_E = 2g_E\Lambda^2 N_\mathcal{B}^d.$
The total cost for HAM-T is then:
\begin{equation}
\T[\text{HAM-T}^\mathcal{B}] = \T[e^{-iH_E^\mathcal{B}}] \log_2 \frac{16(\alpha + \alpha_E)}{\epsilon} + \T[H_{M+GM+B}^\mathcal{B}]\,.
\end{equation}
One also requires an additional $\lceil \log_2 \mathcal{M} \rceil$ qubits.

We then make use of Ref.~\cite[Lemma 6]{Low:2019b} to get the $\T$-gate cost for the simulation of $e^{iH^\mathcal{B}}$ from HAM-T.
In total, we will require $\alpha$ segments of simulation, each of which require one call to $e^{iH_E^\mathcal{B}/(2\alpha)}$ and $3K$ calls to HAM-T where $K = \lceil -1 + \frac{2\ln(2\alpha/\epsilon)}{\ln \ln (2\alpha/\epsilon) + 1}\rceil.$
No additional qubits are required.
The total simulation cost for a single HHKL block to unit time is:
\begin{equation}
\T[e^{iH^\mathcal{B}}] = \alpha \Big(\T[e^{iH_E^\mathcal{B}}] + \lceil -1 + \frac{2\ln(2\alpha/\epsilon)}{\ln \ln (2\alpha/\epsilon) + 1}\rceil \T[\text{HAM-T}^\mathcal{B}]\Big).
\end{equation}

The final cost is then just a multiplicative factor of the above expression:
\begin{equation}
T_{Qubitization} = \frac{N^dT}{N_\mathcal{B}^d}\T[e^{iH^\mathcal{B}}]\,.
\end{equation}
The only remaining unknown parameter is the size of the HHKL blocks, which is determined by the Lieb-Robinson velocity of the system and block simulation time, namely $N_\mathcal{B} = v_{LR}$.
We document our approach to bounding the Lieb-Robinson velocity in Appendix~\ref{app:lr}.

\begin{figure}
    \centering
    \includegraphics[width=\textwidth]{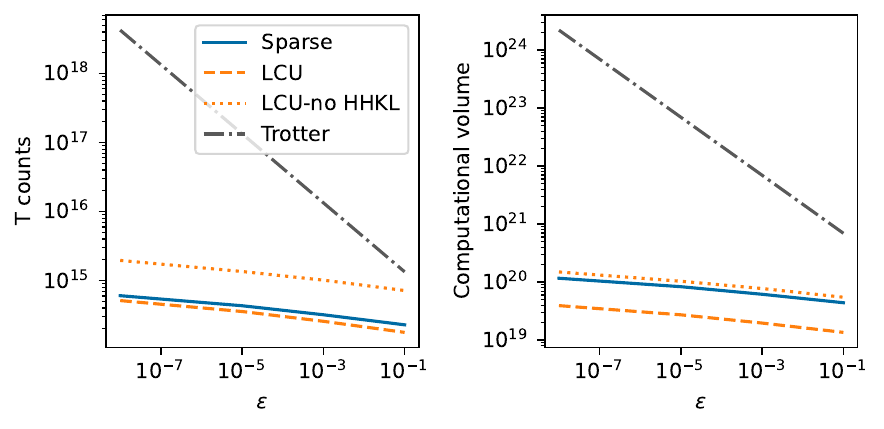}
    \caption{Comparison of resource estimates for Hamiltonian simulation of a two dimensional U(1) lattice gauge theory using the methods in this work, labeled LCU and sparse, and from the Trotterized simulation of Kan and Nam \cite{kan2021} for the case of heavy-ion collision. The left subfigure includes $\T$-count comparisons, and the right the computational volume defined as the $\T$-counts multiplied by the qubit counts.
    Costs are shown for different values of accuracy of the simulation, $\epsilon$.
    Here we take the parameters given by Kan and Nam ($a = 0.1$, $N = 100$, $T = 10$, $\Lambda = 10$) and $m =g = 10$.}
    \label{fig:final-comparison}
\end{figure}

For our direct comparison to Trotterization, we take the physical parameters cited therein Ref.~\cite{kan2021} for the case of simulating heavy-ion collisions on a two dimensional U(1) gauge theory: $a = 0.1$, $N = 100$, $T = 10$, $\Lambda = 10$, and $m = g = 10$.
The Lieb-Robinson velocity and subsequent block-size calculation yields $N_\mathcal{B} \geq  53$.
We consider $\epsilon = 10^{-8},$ as per Kan and Nam, but also include comparison to lower accuracy simulations $\epsilon = 10^{-1}, 10^{-3}, 10^{-5}$ as the accuracy scaling is worse for the Trotterization than for qubitized simulation.
The simulation costs are shown in Fig.~\ref{fig:final-comparison}.

Qubitized simulation outperforms Trotterization for essentially all accuracies, with the cross-over in costs occurring at very low accuracies $\epsilon < 10^{-1}$.
Both qubitized methods perform between two to four orders of magnitude better in $\T$-gate counts, with larger improvements at higher accuracies.
However, we note that there is practically no difference between LCU and sparse implementations of the block encoding: this seems contradictory to our previous results for the block encoding showing $\sim 3\times$ improvements in LCU over sparse block encodings.
The closing of the gap, per se, can be understood by looking at HAM-T, which requires $\log_2(16(\alpha + \alpha_E)/\epsilon) \sim 50$ applications of the electric field term, which becomes comparable in cost to the block encodings themselves.
As such, the improvement in the total cost of implementing HAM-T, the primary query oracle, is diminished as the electric field implementation is agnostic to the block encoding method.

We also report the computational volume, defined as a the $\T$-gate count multiplied by the logical qubit counts.
We see that there are additional improvements here for the qubitized algorithm, where the Trotter requires $5 \times 10^5$ qubits, the sparse and LCU respectively take $1 \times 10^5$ and $5 \times 10^4$ qubits.
Overall, we see six orders of magnitude improvement in computational volume between the Trotterized and qubitized methods.
We have also included a qubitized simulation without using HHKL, where $N_\mathcal{B} = N$.
We find that this na\"ive qubitized simulation outperforms Trotterization, and is slightly worse than HHKL, as expected since we have only 4 HHKL blocks at this parameter range.

For a richer comparison between the techniques, we also compare the resource estimates for different values of other physical parameters.
The qubitized algorithm purports to be optimal in spacetime volume, wherein we may look at different $T$ and $N$.
The case for $T$ is quite straightforward, so we exclude it.
As for $N$, the case is more complex since the real benefit of the qubitized approach is when the HHKL block size is sufficiently small, becoming polylogarithmic in $N$.
It is clear that for the parameters above, this may not be the case, and so we may not be getting the greatest benefits of the technique.

We also vary $g$, $m$, and $a$ as they affect $v_{LR}$, as well as the commutators for Trotterization. 
We find, however, that changing $g$ and $m$ does not significantly affect $v_{LR}$, for example two orders of magnitude variation in either takes $v_{LR} \geq 53 \rightarrow v_{LR} \geq 42$.
The values do change the resource estimates, for example changing $m = g= 10 $ to $m = g = 0.1$ increases the $\T$-gate counts in Trotterization, almost independently of all other parameter variations, by three orders of magnitude.
However, the same applies for qubitization, meaning that the relative improvement in using either method is nearly unchanging with $g$ and $m$.
The remaining parameter is $a$, which highly influences $v_{LR}$ and also the commutators for Trotterization.

We report the required resources in terms of $\T$-gate counts and computational volume, as well as the relative improvement for qubitization over Trotterization in spacetime volume in Table~\ref{tab:final-comparison-matrix} at a fixed $ m = g= 10$, while varying $\epsilon \in \{10^{-3}, 10^{-1}\}$, $N \in \{100, 1000 \}$, $a \in \{10^{-2}, 10^{-1}, 1\}$.
For sake of readability, we have only included the LCU results here.
We see a general trend, namely that increasing $a$ is the primary improvement, indicating that Trotterization is preferred for small $a$.
Further, as $\epsilon$ decreases, improvements increase, indicating that at high accurracies the qubitized simulation outperforms Trotterization.

We note, however, that the expected improvements in qubitization with respect to $N$ are not seen,
and an opposite trend is observed.
Second order Trotterization scales as
$\mathcal{O}(N^{3/2})$, na\"ive qubitization as $\mathcal{O}(N^2)$, and HHKL as $\mathcal{O}(N)$, up to logarithmic factors.
For qubitization, the $\mathcal{O}(N^2)$ scaling is a product of the $\mathcal{O}(N)$ Hamiltonian norm appearing as a factor in query complexity and the $\mathcal{O}(N)$ block encoding cost.
Therefore, the HHKL scaling is only $\mathcal{O}(N)$, \emph{asymptotically}, in the regime where the sizes of blocks simulated via qubitization are much smaller than the system size, which is \textit{not} the case for U(1).
As such, we see that the U(1) scaling is closer to the na\"ive qubitization scaling, and hence scales worse with $N$ than Trotterization.
If one were able to find tighter bounds for $v_{LR}$, this may not be the case.

As it stands, we find that the Trotterization scheme of Kan and Nam~\cite{kan2021} will outperform our qubitized implementation for simulations with asymptotically small $a$, large $\epsilon$, and large $N$.
Physically, $a$ dictates the length scale of physical interest and $N$ the system size.
While we do not comment on the quantitative regimes where qubitization outperforms Trotterization, we note that resources pertaining to a particular simulation task can be computed from our resource estimate expressions.
A quantitative analysis would be of particular interest to U(1) gauge theory since in 3+1D and 2+1D, U(1) has massless photon interactions, for which finite size corrections scale only polynomially with system size \cite{PhysRevX.9.021022}, in contrast to SU(N) theories which scale exponentially.
As such, the quantitative assessment of "small $a$, large $N$" would be particularly important in determining the benefit of qubitization over Trotterization for U(1) LGTs.

\begin{table}[]
    \centering
    \begin{tabular}{|c|c|c|c|c|c|c|c|c|}
\hline
          $\epsilon$ &                $N$ &                 $a$ &              $v_{LR}$ &          $T_\textrm{Trotter}$ &        $T_\textrm{Qubit.}$ &         $Q_\textrm{Trotter}$ &  $Q_\textrm{Qubit.}$    & Improvement \\
\hline
$10^{-3}$ & $10^{2}$ &  $10^{0\hphantom{-}}$ & $1.0\times10^{1}$ & $3.2	\times10^{15}$ & $4.7	\times10^{12}$ & $5.1	\times10^{5}$ & $8.1	\times10^{4}$ & $4.4	\times10^{3}$ \\
 & & $10^{-1}$ & $5.3	\times10^{1}$ & $1.3	\times10^{16}$ & $2.6	\times10^{14}$ & $5.1	\times10^{5}$ & $7.6	\times10^{4}$ & $3.5	\times10^{2}$ \\
 &  & $10^{-2}$ & $4.2	\times10^{2}$ & $1.1	\times10^{17}$ & $7.8	\times10^{15}$ & $5.1	\times10^{5}$ & $7.7	\times10^{4}$ & $9.8	\times10^{1}$ \\
 \hline
 & $10^{3}$ &  $10^{0\hphantom{-}}$ & $1.0\times10^{1}$ & $3.2	\times10^{18}$ & $4.7	\times10^{14}$ & $5.1	\times10^{7}$ & $8.1	\times10^{6}$ & $4.4	\times10^{4}$ \\
&  & $10^{-1}$ & $5.3	\times10^{1}$ & $1.3	\times10^{19}$ & $2.6	\times10^{16}$ & $5.1	\times10^{7}$ & $7.7	\times10^{6}$ & $3.5	\times10^{3}$ \\
 & & $10^{-2}$ & $4.2	\times10^{2}$ & $1.1	\times10^{20}$ & $1.7	\times10^{19}$ & $5.1	\times10^{7}$ & $7.6	\times10^{6}$ & $4.4	\times10^{1}$ \\
 \hline
$10^{-1}$ & $10^{2}$ &  $10^{0\hphantom{-}}$ & $1.0\times10^{1}$ & $3.2	\times10^{14}$ & $3.1	\times10^{12}$ & $5.1	\times10^{5}$ & $8.1	\times10^{4}$ & $6.9	\times10^{2}$ \\
 & & $10^{-1}$ & $5.3	\times10^{1}$ & $1.3	\times10^{15}$ & $1.8	\times10^{14}$ & $5.1	\times10^{5}$ & $7.7	\times10^{4}$ & $5.1	\times10^{1}$ \\
 &  & $10^{-2}$ & $4.2	\times10^{2}$ & $1.1	\times10^{16}$ & $5.6	\times10^{15}$ & $5.1	\times10^{5}$ & $7.7	\times10^{4}$ & $1.4	\times10^{1}$ \\
 \hline
 & $10^{3}$ &  $10^{0\hphantom{-}}$ & $1.0\times10^{1}$ & $3.2	\times10^{17}$ & $3.1	\times10^{14}$ & $5.1	\times10^{7}$ & $8.1	\times10^{6}$ & $6.9	\times10^{3}$ \\
 & & $10^{-1}$ & $5.3	\times10^{1}$ & $1.3	\times10^{18}$ & $1.8	\times10^{16}$ & $5.1	\times10^{7}$ & $7.7	\times10^{6}$ & $5.1	\times10^{2}$ \\
 & & $10^{-2}$ & $4.2	\times10^{2}$ & $1.1	\times10^{19}$ & $1.3	\times10^{19}$ & $5.1	\times10^{7}$ & $7.6	\times10^{6}$ & $6.0	\times10^{0}$ \\
\hline
\end{tabular}

    \caption{Comparison of Trotterization and HHKL based simulation using an LCU block encoding for U(1) lattice gauge theory in two dimensions for a heavy ion simulation. Various accuracies $\epsilon$, linear system sizes $N$, and lattice spacings $a$ are shown. 
    Bounds on the Lieb-Robinson velocity $v_{LR}$, $\T$-gate and qubit counts are derived therein, and a final Improvement in spacetime volume of the computation ($\T$-counts times logical qubit counts) is reported.}
    \label{tab:final-comparison-matrix}
\end{table}

\section{Generalization to non-Abelian gauge theories and higher dimensions \label{sec:5}}
The explicit circuits for block encoding the Kogut-Susskind Hamiltonian we have provided can be readily adapted to higher spacetime dimensions and higher order non-Abelian LGTs, such as SU(2) and SU(3). While the VC encoding~\cite{cirac2005} is sufficient for U(1) LGTs in two dimensions, it is more resource efficient to consider the GS~\cite{Setia:2019a} encoding for SU(2) and SU(3) in higher dimensions. The advantage of the GS encoding is that it is easily amenable to changing lattice geometry, number of colors, and spatial dimension and uses only one extra qubit per lattice site, in contrast to the $\mathcal{O}(n_c)$ extra qubits required in the VC encoding. Additionally, its structure and ordering follow closely to that of the VC encoding, so the circuit constructions we employ will follow the same logic.

Using the GS construction, we already saw that considering a $d$ dimensional cubic lattice on $N^d$ sites for simulating an SU($n_c$) LGT will require $(d+n_c-1)N_\mathcal{B}^d$ qubits to store the fermionic degrees of freedom. Unlike fermions, the bosonic interactions are always local, so there is no need to consider any special form of encoding. Instead, the bosonic encoding will scale directly in terms of the spatial dimension and the specific LGT. For $d$ spatial dimensions, each lattice site will store $d$ bosons, corresponding to the $d$ ordered edges, and for SU($n_c$), each bosonic state will have $n_c^2-1$ registers, each of size at most $\log\Lambda$, indexing the local quantum numbers. Thus, the total number of qubits needed to store the wavefunction is
\begin{equation}
    Q=(d+n_c-1)N^d+ dN^d(n_c^2-1)\log\Lambda\,.
\end{equation}

Rather than providing full resource estimates for both the sparse and LCU encoding models as for the U(1) case, here we restrict ourselves to providing resource requirements in only the sparse access model for three reasons. First, as noted in the U(1) case, the difference in resource requirements between the sparse and LCU models for full simulation is insignificant with respect to the dominant electric field term, making either encoding model a valid candidate. Second, the gauge field operators $U_{ab}(\vec{n},\hat{l})$ are conventionally written in a form suitable to the sparse access model (see Eqs.~\eqref{eq:su2-magnetic} and~\eqref{eq:su3_gauge}). In the LCU model one must determine a sequence of elementary gates to implement these operators directly on basis states, which is not only non-trivial, but will be unique to each LGT considered, and thus does not generalize well to SU($n_c$). We remark that a possible implementation of the gauge field operator could be realized via a modification to the Clebsch-Gordan transform~\cite{Bacon:2006a}. One would need to adjust the quantum circuit to transform both the left and right ends of a lattice link and include the dimensional normalization prefactor. Additionally, the generalization of the Clebsch-Gordan transform to SU($n_c$) relies on using qudits, so one would need to construct the Clebsh-Gordan transform for SU(3) on qubits. Third, provided one could find an implementation for \textsc{select}, then the resource requirements could be determined straightforwardly from the expressions given in the U(1) case, as the complexity of the unary iteration subroutine will scale directly with the spatial dimension and number of colors in the LGT. One would only need to account for the cost of implementing the controlled-$U_{ab}$ gates. 

The logic of our algorithm remains relatively unchanged, with only modifications to the operators being applied in the corresponding LGT. The mass term straightforwardly generalizes, scaling with the number of sites and colors. In the sparse access model, only $O_H$ contributes to the complexity scaling, which  requires a controlled-SID gate for each lattice site and each color DOF. Additionally, the size of the registers being incremented or decremented will scale with the type of LGT. Thus, for SU($n_c$) in $d$ spatial dimensions, we have
\begin{equation}
    \T[O_H(H_M)]=4n_cN^d(2\log (N^d+n_c-2)+1)\,.
\end{equation}

The remaining Hamiltonian terms all have contributions from the gauge field operators, which will more strictly depend on the specific LGT being considered. At this point, we will restrict to SU(2) and SU(3) LGTs. The simplest of the gauge terms is the electric field term, which again will be fast-forwarded. In SU(2), the Casimir operator is given by Eq.~\eqref{eq:su2-electric} and in SU(3) by Eq.~\eqref{eq:su3_electric}.

The cost of implementing these operators can be determined again using the circuit for fast-forward evolution in Fig.~\ref{fig:fast-forward circuit}. For SU(2), either the QROM or Karatsuba integer multiplication can be used, as before, to implement a unitary $U_k$ which prepares the eigenvalues of the Casimir operators, depending on the bosonic truncation parameter $\Lambda$. Here, we provide resource estimates using a QROM as it outperforms integer multiplication for physically relevant values of $\Lambda$.

On the other hand, for SU(3), there are now two representation indices characterizing the action of the Casimir operator meaning a QROM will scale as $\mathcal{O}(\Lambda^2)$. The alternative approach of implementing the integer arithmetic on-the-fly will also require more work. In addition to the Karatsuba multiplication algorithm~\cite{Parent:2017a}, we also require algorithms for integer addition~\cite{Gidney:2018a} and integer division~\cite{Thapliyal:2017a}. There is again a tradeoff in optimal complexity depending on $\Lambda$. Numerically, for $\Lambda\lesssim 90$, the QROM is more efficient. The associated complexities are then given by
\begin{align}
    &\hphantom{\text{(Arithm.) } \text{SU(3): } \T[e^{-itH_E^\mathcal{B}}]}
    \mathllap{\text{SU(2): } \T[e^{-itH_E^\mathcal{B}}]}
    =2(4\Lambda-4)+4\log(\Lambda(\Lambda+1))\log(1/\epsilon)\,,
    \\
    &\hphantom{\text{(Arithm.) } \text{SU(3): } \T[e^{-itH_E^\mathcal{B}}]}
    \mathllap{\text{(QROM) } \text{SU(3): } \T[e^{-itH_E^\mathcal{B}}]}
    =2(2\Lambda^2-4)+4\log(\Lambda(\Lambda+3))\log(1/\epsilon)\,, \\
    &\begin{alignedat}{9}
        \text{(Arithm.) } \text{SU(3): } \T[e^{-itH_E^\mathcal{B}}]=2(99\log^2(\Lambda(\Lambda+3))&-60\log(\Lambda(\Lambda+3))-16)
        \\
        & +4\log(\Lambda(\Lambda+3))\log(1/\epsilon)]\,.
    \end{alignedat}
\end{align}

 The oracle constructions for the gauge field operators $U_{ab}(\vec{n},\hat{l})$ are also required for both the gauge-matter and magnetic field interaction terms. For SU(2) the gauge field operators are given in Eq.~\eqref{eq:su2-magnetic} and for SU(3) in Eq.~\eqref{eq:su3_gauge}. 

In the sparse access model, we generalize the bosonic part of the circuit constructions for $O_F$ and $O_H$ in Figs.~\ref{fig:vert_gm_sparse_enum} and~\ref{fig:vert_gm_sparse_value}. The fermionic part of the gauge-matter interaction will follow the same logic as in the U(1) case. A QROM will load in the binary symplectic vectors associated with the particular Pauli representation chosen. Then the non-zero final states are given by $|f\oplus a_\ell\rangle$, just as before. The only difference is that now, in addition to their being a unique symplectic vector for every site, there will also be a unique vector for every color configuration, meaning the QROM contains $n_c^2N_\mathcal{B}^d$ elements in the database.

In order to implement the gauge field operators, though, we must be a bit more careful than in the U(1) case. For the U(1) LGTs we were able to implement the gauge field operators directly onto the bosonic registers, as they corresponded to increments or decrements of occupancies of individual bosonic modes. On the other hand, for SU(2) and SU(3), the gauge-field operators require preparing a superposition of final states.

For our particular method we make use of a space-time tradeoff. Rather than storing a superposition of final states on a single register, we store each final state in its own register, increasing the spatial resource requirements by a constant prefactor overhead. The number of additional registers will depend on the specific LGT, but for SU(2), one additional register is required and for SU(3) (in the worst case) eleven additional registers are required, each of size $O(\log\Lambda)$. With this approach, computing the final states amounts to performing only controlled-increments and controlled-decrements on the state indices. Additionally, with each non-zero final state stored in its own register, then the unique matrix element values for each state can be computed in $O_H$ on these separate registers.

While we make use of the approach involving additional state registers for computing resource requirements, we also provide some alternative, less intuitive, approaches one could employ. First, rather than storing each entire final state, we can include an $\mathcal{O}(1)$ size register of indicator qubits which determine the final state we are in. In SU(2), for example, the action of the gauge field operator would map $\ket{j}\mapsto \{\ket{j}\ket{0},\ket{j}\ket{1}\}$ where the state of the indicator qubit corresponds to the states $\ket{j\pm 1/2}$. However, since computing the matrix element values in $O_H$ requires arithmetic, the explicit circuit construction would require additional controlled operations on the indicator qubit to prepare the correct binary represenation of the state prior to doing the arithmetic. Another alternative approach applicable only to SU(2) involves decomposing the gauge field operators into sums of one-sparse terms, which are easy to encode via the sparse access oracles~\cite{kirby2020variational}. For the magnetic field term $H_B$, though, this amounts to exchanging the sums of the gauge-field operators and the trace operation over the plaquettes, which is only possible in SU(2), as the isospin parameters in SU(3) are color-dependent.

Using the approach to store each entire final state, for the bosonic part in SU(2) the non-zero final states $|J,M_L,M_R\rangle$ are constructed via a series of increments or decrements, similar to the U(1) case. However, for SU(2), we now need six increments or decrements for each gauge field operator with fixed color indices, three for each of the $|j\pm 1/2\rangle$ final states. Thus, in Fig.~\ref{fig:vert_gm_sparse_enum}, we can replace each controlled-SID gate with six controlled-SID gates (three to each of the two registers storing the two final states). The rest of the logic will follow exactly as in the U(1) case.

For SU(3), the generalization is similar. The isospin and hypercharges will be incremented or decremented according to the fixed color indices. However, unlike SU(2), the SU(3) gauge field operators have sums over $T_L'$ and $T_R'$ in addition to the sum over the representation indices. While $T_L'$ and $T_R'$ are color-dependent, $t_L,t_R\in\{0,1/2\}$, so in the worst case, we obtain at most four additional non-zero final states. Thus, there are a total of twelve non-zero final states, each one requiring either seven or eight controlled-SID gates. So each controlled-decrement gate in Fig.~\ref{fig:vert_gm_sparse_enum} will now be replaced with 88 controlled-SID gates. The resulting complexities for the gauge-matter and magnetic field term then become
\begin{align}
    \text{SU(2): }\T[O_F(H_{GM})]&=4(4N^d)-4+96N^d(2\log\Lambda+4)\,, \\
    \text{SU(3): }\T[O_F(H_{GM})]&=4(9N^d)-4+3168N^d(2\log\Lambda+4)\,.
\end{align}

While the implementation of $O_H$ was trivial for the gauge fields in U(1), it is not so for SU(2) and SU(3) and we must include additional subroutines to the circuit in Fig.~\ref{fig:vert_gm_sparse_value} to obtain the bosonic matrix element values. In both SU(2) and SU(3), this amounts to computing the Clebsch-Gordan coefficients and the dimension normalization prefactor. While each of these amplitudes is known \emph{a priori}, allowing us to use a QROM to classically pre-compute and load these values in, we will see that it is in fact more efficient to do the arithmetic on-the-fly. For SU(2), there are four Clebsch-Gordan coefficients $c_{\Delta j,\Delta m}$ indexed by the change in angular momentum $\Delta j=J-j$ and change in the $z$-component of angular momentum $\Delta m=M-m$. However, each of these coefficients is a function of $j$ and $m$. Since $j\in[0,\Lambda]$ (with half-integer steps) and $m\in[-\Lambda,\Lambda]$, there are $16\Lambda^2$ elements in the classical database. On the other hand, each of the Clebsch-Gordan coefficients is a simple function of $j$ and $m$, and can be readily computed using the integer arithmetic subroutines from the electric field term in addition to a subroutine for computing the square root~\cite{Sultana:2011a}. The dimension normalization prefactor can also be readily computed in this way.

For SU(3), the same logic applies, though now the arithmetic is a bit more involved, but still more efficient than implementing a QROM for a database of size $\mathcal{O}(\Lambda^5)$. The SU(3) Clebsch-Gordan coefficients can be computed from the product of the isoscalar factors $I_{ab}$ and the SU(2) Clebsch-Gordan coefficients with $\Delta j$ and $\Delta m$ replaced by $\Delta T$ and $\Delta T^z$. In contrast to SU(2), there are now fifteen Clebsch-Gordan coefficients, each of which is a function of the representation indices $p$ and $q$, isospin $T$, hypercharge $Y$, and $z$-component of the isospin $T^z$. For a more detailed review, see Ref.~\cite[Appendix D]{kan2021}.

Including these additional costs in to the circuits of Fig.~\ref{fig:vert_gm_sparse_value}, we obtain the resource requirements for the gauge-matter term of the Hamiltonian in both SU(2) and SU(3):
\begin{align}
    \text{SU(2): }\T[O_H(H_{GM})]&=4(4(4N^d)-4)+3+8(d+1)N^d+2(71\log^2\Lambda-16\log\Lambda-32)\,,\\
    \text{SU(3): }\T[O_H(H_{GM})]&=4(4(9N^d)-4)+3+8(d+2)N^d+12(263\log^2\Lambda-28\log\Lambda-164)\,.
\end{align}

Finally, the magnetic term is realized as the product of four gauge field operators around a plaquette in the lattice with the additional constraint that the color indices contract to singlets at each vertex, denoted by the trace operation. Thus the non-zero final states will be given by the tensor product of incremented/decremented basis states resulting from the action of the gauge field operators. The complexity is four times as costly as implementing the encoding oracle for a single gauge-field term, and we can use unary iteration to index over the lattice plaquettes. The number of plaquettes in a $d$-dimensional hypercubic lattice is
\begin{equation}
    N_\Box=\binom{d}{2}N^2(N+1)^{d-2}\,.
\end{equation}
The associated complexities are
\begin{align}
    \text{SU(2): }\T[O_F(H_{B})]&=4(2^4N_\Box)-4+384N^d(2\log\Lambda+1)\,,\\\
    \text{SU(2): }\T[O_H(H_{B})]&=8(71\log^2\Lambda-16\log\Lambda-32)\,,\\
    \text{SU(3): }\T[O_F(H_{B})]&=4(3^4N_\Box)-4+12\,672N^d(2\log\Lambda+1)\,, \\
    \text{SU(3): }\T[O_H(H_{B})]&=48(263\log^2\Lambda-28\log\Lambda-164)\,.
\end{align}

With each of the oracle complexities, we can give the full block encoding complexities for each Hamiltonian term. Since we are actually interested in a block encoding of $H_M+H_{GM}+H_B$, we need two additional levels of control on each individual encoding circuit as well as two single qubit rotations to implement an LCU encoding of our block encodings. The doubly-controlled block encoding complexities for SU(2) are
\begin{align}
    &\hphantom{\T[CC-H_{GM}]}
    \mathllap{\T[CC-H_M]}
    =32N^d+16N^d(2\log(N^d)+5)+16\log(N^d)\log(1/\epsilon)\,, \\
    \label{eq:cchgmsu2}
    &\begin{alignedat}{9}
    \T[CC-H_{GM}]&=48(d+1)N^d+2\big[4(4N^d)+4+96N^d(2\log\Lambda+8)+4(4(4N^d)-4)\\  &+ 5+24(d+1)N^d+2(684\log^2\Lambda-432\log\Lambda+16)\big]+32p\log(1/\epsilon)\,,    
    \end{alignedat}
    \\
    &\begin{alignedat}{9}
    \hphantom{\T[CC-H_{GM}]}
    \mathllap{\T[CC-H_B]}
    &=2\bigl[4(2^4N_\Box)+4+384N^d(2\log\Lambda+1)
    \\
    &+8(684\log^2\Lambda-432\log\Lambda+16)\bigr]+16p\log(1/\epsilon)\,,    
    \end{alignedat}
\end{align}
and SU(3) are
\begin{align}
    &\hphantom{\T[CC-H_{GM}]}
    \mathllap{\T[CC-H_M]}
    =48N^d+24N^d(2\log(N^d+1)+5)+16\log(N^d+1)\log(1/\epsilon)\,, \\
     \label{eq:cchgmsu3}
     &\begin{alignedat}{9}
     \T[CC-H_{GM}]&=48(d+2)N^d+2\big[4(4N^d)+4+12\,672N^d(2\log\Lambda+8)
     \\
     &+ 4(4(9N^d)-4) +5+24(d+2)N^d
     \\
     &+12(2\,988\log^2\Lambda-2\,028\log\Lambda+148)\big]+32p\log(1/\epsilon)\,,
     \end{alignedat}
    \\
    &\begin{alignedat}{9}
    \hphantom{\T[CC-H_{GM}]}
    \mathllap{\T[CC-H_B]}
    &=2\bigl[4(3^4N_\Box)+4+384N^d(2\log\Lambda+1)
    \\&+48(2\,988\log^2\Lambda-2\,028\log\Lambda+148)\bigr]+16p\log(1/\epsilon)\,,    
    \end{alignedat}
\end{align}
where $p = 64$ corresponds to double-precision accuracy for the matrix element values.

\begin{figure}[ht]
        \centering
    \includegraphics[width=\textwidth]{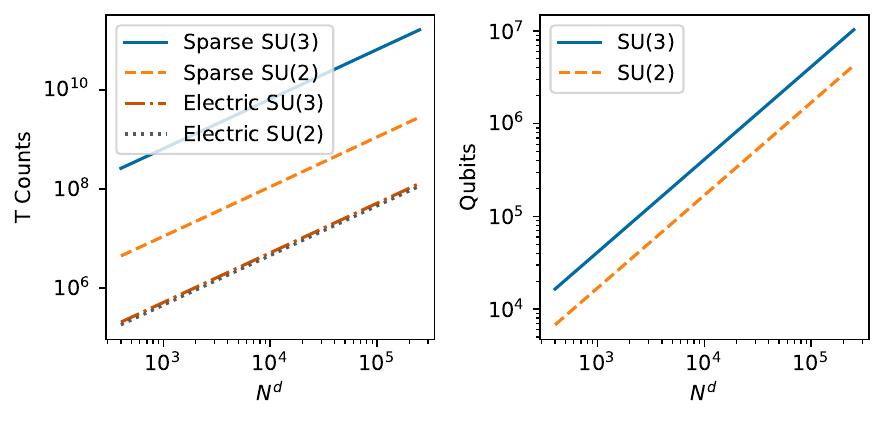}
    \caption{\label{fig:total_counts_su2} The total (a) $\T$-counts and (b) qubit counts needed to block encode the Hamiltonian for SU(2)  and SU(3) gauge theories.
    Also shown are the electric field costs, which are necessary to compute the cost of HAM-T.}
\end{figure}

In Fig.~\ref{fig:total_counts_su2}, we show a comparison of the different components of HAM-T, similar to Fig.~\ref{fig:total_counts}, for SU(2) and SU(3).
We see that for SU(2), the block encoding cost is about 25x the cost of the electric field term, in concordance with the U(1) calculations.
This means that for HAM-T, the electric field term still dominates, and hence the method of block encoding is not particularly important.
Proceeding with only the sparse encoding will still yield good resource estimates.
As for SU(3), the ratio is closer to 1000, and so the block encoding is finally the dominant cost in HAM-T.
Accordingly, improvements in block encoding techniques, such as LCU, may yield improvements in SU(3) simulation relative to the numbers we will report with our sparse encodings.

\begin{table}[]
    \centering
    \begin{tabular}{|c|c|c|c|c|c|c|c|}
\hline
          $\epsilon$ &                $N$ &                 $a$ &              $T_\textrm{Trotter}$ &        $T_\textrm{Qubit.}$ &         $Q_\textrm{Trotter}$ &  $Q_\textrm{Qubit.}$    & Improvement \\
\hline
$10^{-3}$ & $10^{3}$ &  $10^{0}$ & $1.2	\times10^{36}$ & $1.4	\times10^{26}$ & $10^{11}$ & $3.4	\times10^{10}$ & $2.9	\times10^{10}$ \\
 &  & $10^{-1}$ & $3.9	\times10^{37}$ & $2.2	\times10^{26}$ & $10^{11}$ & $2.4	\times10^{10}$ & $7.1	\times10^{11}$ \\
 &  & $10^{-2}$ & $1.2	\times10^{39}$ & $9.6	\times10^{26}$ & $10^{11}$ & $2.4	\times10^{10}$ & $5.1	\times10^{12}$ \\
 & $10^{2}$ &  $10^{0}$ & $3.9	\times10^{31}$ & $1.0	\times10^{20}$ &  $10^{8}$ &  $2.4	\times10^{7}$ & $1.5	\times10^{12}$ \\
& & $10^{-1}$ & $1.2	\times10^{33}$ & $1.5	\times10^{20}$ &  $10^{8}$ &  $2.4	\times10^{7}$ & $3.1	\times10^{13}$ \\
 & & $10^{-2}$ & $3.9	\times10^{34}$ & $6.9	\times10^{20}$ &  $10^{8}$ &  $2.4	\times10^{7}$ & $2.2	\times10^{14}$ \\
$10^{-1}$ & $10^{3}$ &  $10^{0}$ & $1.2	\times10^{35}$ & $6.8	\times10^{25}$ & $10^{11}$ & $2.4	\times10^{10}$ &  $7.3	\times10^{9}$ \\
& & $10^{-1}$ & $3.9	\times10^{36}$ & $1.0	\times10^{26}$ & $10^{11}$ & $2.4	\times10^{10}$ & $1.5	\times10^{11}$ \\
& & $10^{-2}$ & $1.2	\times10^{39}$ & $4.4	\times10^{26}$ & $10^{11}$ & $2.4	\times10^{10}$ & $1.1	\times10^{13}$ \\
 & $10^{2}$ &  $10^{0}$ & $3.9	\times10^{30}$ & $4.7	\times10^{19}$ &  $10^{8}$ &  $2.4	\times10^{7}$ & $3.3	\times10^{11}$ \\
 & & $10^{-1}$ & $1.2	\times10^{32}$ & $7.2	\times10^{19}$ &  $10^{8}$ &  $2.4	\times10^{7}$ & $6.9	\times10^{12}$ \\
&  & $10^{-2}$ & $3.9	\times10^{33}$ & $3.2	\times10^{20}$ &  $10^{8}$ &  $2.4	\times10^{7}$ & $4.9	\times10^{13}$ \\
\hline
\end{tabular}

    \caption{Comparison of Trotterization and qubitized simulation using a sparse block encoding for SU(2) lattice gauge theory in three dimensions for a heavy ion simulation. Various accuracies $\epsilon$, linear system sizes $N$, and lattice spacings $a$ are shown. 
    $\T$-gate and qubit counts are derived therein, and a final Improvement in spacetime volume of the computation ($\T$-counts times logical qubit counts) is reported. Note that we use na\"ive qubitization here, with no HHKL decomposition.}
    \label{tab:final-comparison-matrix-su2}
\end{table}

\begin{table}[]
    \centering
    \begin{tabular}{|c|c|c|c|c|c|c|c|}
\hline
          $\epsilon$ &                $N$ &                 $a$ &              $T_\textrm{Trotter}$ &        $T_\textrm{Qubit.}$ &         $Q_\textrm{Trotter}$ &  $Q_\textrm{Qubit.}$    & Improvement \\
\hline
$10^{-3}$ & $10^{3}$ &  $10^{0\hphantom{-}}$ & $1.0\times10^{50}$ & $2.0	\times10^{27}$ & $2.6	\times10^{11}$ & $6.0	\times10^{10}$ & $2.1	\times10^{23}$ \\
 &  & $10^{-1}$ & $3.2	\times10^{51}$ & $2.9	\times10^{27}$ & $2.6	\times10^{11}$ & $6.0	\times10^{10}$ & $4.6	\times10^{24}$ \\
& & $10^{-2}$ & $1.0\times10^{53}$ & $1.2	\times10^{28}$ & $2.6	\times10^{11}$ & $6.0	\times10^{10}$ & $3.5	\times10^{25}$ \\
 & $10^{2}$ &  $10^{0\hphantom{-}}$ & $3.2	\times10^{45}$ & $1.6	\times10^{21}$ &  $2.6	\times10^{8}$ &  $6.0	\times10^{7}$ & $8.4	\times10^{24}$ \\
 & & $10^{-1}$ & $1.0\times10^{47}$ & $2.3	\times10^{21}$ &  $2.6	\times10^{8}$ &  $6.0	\times10^{7}$ & $1.8	\times10^{26}$ \\
&  & $10^{-2}$ & $3.2	\times10^{48}$ & $1.0\times10^{22}$ &  $2.6	\times10^{8}$ &  $6.0	\times10^{7}$ & $1.3	\times10^{27}$ \\
$10^{-1}$ & $10^{3}$ &  $10^{0\hphantom{-}}$ & $1.0\times10^{49}$ & $1.7	\times10^{27}$ & $2.6	\times10^{11}$ & $6.0	\times10^{10}$ & $2.5	\times10^{22}$ \\
 & & $10^{-1}$ & $3.2	\times10^{50}$ & $2.5	\times10^{27}$ & $2.6	\times10^{11}$ & $6.0	\times10^{10}$ & $5.4	\times10^{23}$ \\
 & & $10^{-2}$ & $1.0\times10^{52}$ & $1.0\times10^{28}$ & $2.6	\times10^{11}$ & $6.0	\times10^{10}$ & $4.1	\times10^{24}$ \\
 & $10^{2}$ &  $10^{0\hphantom{-}}$ & $3.2	\times10^{44}$ & $1.3	\times10^{21}$ &  $2.6	\times10^{8}$ &  $6.0	\times10^{7}$ & $1.0\times10^{24}$ \\
 &  & $10^{-1}$ & $1.0\times10^{46}$ & $1.9	\times10^{21}$ &  $2.6	\times10^{8}$ &  $6.0	\times10^{7}$ & $2.1	\times10^{25}$ \\
 & & $10^{-2}$ & $3.2	\times10^{47}$ & $8.4	\times10^{21}$ &  $2.6	\times10^{8}$ &  $6.0	\times10^{7}$ & $1.6	\times10^{26}$ \\
\hline
\end{tabular}

    \caption{Comparison of Trotterization and qubitized simulation using a sparse block encoding for SU(3) lattice gauge theory in three dimensions for a heavy ion simulation. Various accuracies $\epsilon$, linear system sizes $N$, and lattice spacings $a$ are shown. 
    $\T$-gate and qubit counts are derived therein, and a final Improvement in spacetime volume of the computation ($\T$-counts times logical qubit counts) is reported. Note that we use na\"ive qubitization here, with no HHKL decomposition.}
    \label{tab:final-comparison-matrix-su3}
\end{table}

In Table~\ref{tab:final-comparison-matrix-su2} and Table~\ref{tab:final-comparison-matrix-su3}, we provide resource comparisons for a three dimensional simulation of SU(2) and SU(3) LGTs between our qubitized simulation, without HHKL and using the sparse block encoding, to the Trotterized implementation of Kan and Nam~\cite{kan2021}.
We find for SU(2) at minimum 10 orders of magnitude improvement in computational volume, and for SU(3) with a minimum of 23 orders of magnitude improvement.

We remark that the main contribution to the large improvement over the results of Ref.~\cite{kan2021} is a result of the particular representation used for the gauge field operators $U_{ab}$. Kan and Nam map the explicit gauge-field terms of Eqs.~\eqref{eq:su2-magnetic} and~\eqref{eq:su3_gauge} to a block-diagonal representation of the operator which consists of $\sim 10^{12}$ terms for SU(2) and $\sim 10^{21}$ terms for SU(3). This operation dominates their complexity, as the number of terms dictates the number of calls to their arithmetic oracle, but allows them to exponentiate the gauge field operators for Trotterization. We do not encounter this problem using the sparse access block encoding model because the sparse access oracles do not require any structure on the individual Hamiltonian terms, so we do not need a representation in which individual terms are Hermitian or unitary. As discussed in Sec.~\ref{sec:trotterization} and Sec.~\ref{sec:qubitization} the overhead of the block-diagonalization procedure of Ref.~\cite{kan2021} is largely related to the exponential scaling with respect to $n_c$ while the sparse encoding model only has a polynomial dependence on $n_c$.

\section{Conclusion and outlook on the simulation of LGTs \label{sec:6}}
In this work, we have a carried out a careful analysis of qubitized simulation of LGTs.
We build on and fully develop the qubitized Hamiltonian simulation algorithm proposed by Tong \textit{et al.} \cite{Tong:2022a} for U(1), SU(2) and SU(3) gauge theories with matter fields in arbitrary spatial dimension.
In doing so, we worked out the details of various locality-preserving mappings of these models onto qubits~\cite{cirac2005,Setia:2019a,whitsuper}, constructed block encodings for the corresponding qubit Hamiltonians based on sparse and Linear Combination of Unitaries (LCU) oracles, and used the block encodings to construct a full front-to-back analysis of the qubitized simulation algorithm, including estimates of Lieb-Robinson velocities for U(1) gauge theory and a novel signed-increment-decrement circuit.

Additionally, we carried out resource estimates ($\T$-gate counts and qubit counts) for the entire simulation cost within the interaction picture, and compare our numbers to those obtained in Ref.~\cite{kan2021} wherein Trotterized time evolution was considered.
For the case of simulating the SU(3) model in three spatial dimensions our algorithm leads to 25 orders of magnitude lower computational volume requirements compared to Trotterization.
A key ingredient for reaching such an improvement was leveraging the sparsity of color gauge field operators, which is possible in both the sparse and LCU models of block encoding.
The resulting algorithms have polynomial dependence on the number of colors $n_c$, an exponential improvement over state-of-the-art Trotterized simulation techniques \cite{Davoudi:2022xmb, kan2021}.

Given the significant improvements found in this work, one may wonder what additional improvements can be made, and what the results mean in the context of other approaches to the simulation of LGTs on quantum devices.
In terms of additional improvements to the algorithm in this work, we note that the dominant contribution to the cost of the algorithm for small gauge groups like U(1) and SU(2) come from the fast-forwarding of the diagonal electric field term.
Given its asymptotically optimal scaling with the Hamiltonian norm, significant improvements in this direction are unlikely.
However, in the case of SU(3) gauge theories, and more complex gauge groups which arise in grand unification theories such as SU(5)~\cite{Georgi:1974sy} or SO(10)~\cite{Fritzsch:1974nn}, the sparse block-encoding of the magnetic plaquette term is the dominant cost.
It is likely that progress can be made here in reducing resource requirements by making use of LCU implementations utilizing efficient circuits for Schur and Clebsch-Gordan transformations \cite{Bacon:2006a}.

We note, however, that the efficient Trotterized simulation of LGTs is by no means a lost cause.
While a na\"ive look at our reported improvements for SU(3) would indicate that Trotterization would always fall behind qubitization, independent of system size, error, and lattice spacing, the overheads for Trotterization come mainly from a costly decomposition of the plaquette term in Ref.~\cite{kan2021} and the expensive arithmetic therein, required for maintaining gauge invariance.
Recent work~\cite{Gustafson:2022a,Gustafson:2023a,Gustafson:2024a,Lamm:2024a} indicates that these costs can be radically improved with two steps: first, working in the magnetic field basis, and secondly, discretizing the gauge group, e.g., going from U(1) to $\mathbb{Z}_N$.
In doing so, these methods can implement the plaquette term in Trotterization while maintaining gauge invariance without expensive arithmetic.
Even with these improvements, Trotterization would still be asymptotically worse than qubitized simulation, but extrapolating from electronic structure simulation results \cite{Rubin:2023a}, it is likely that Trotterization with these improvements would beat qubitization for small $N$ and large $\epsilon$.

Trotterization in this regime has another distinct advantage, namely that one may be able to numerically compute tight, and perhaps even state-dependent, Trotterization bounds~\cite{ahinolu2021, Yi2022}.
It is therefore likely that such Trotterized simulation methods would outpace qubitization for small simulations with low accuracy requirements.
On the other hand, if one required high accuracies and large simulation cells, and in particular if one were unwilling to incur errors from discretizing the gauge group, the qubitized simulation would certainly be better.

Other recent developments in the simulation of lattice gauge theories can be used to improve both the qubitized and Trotterized simulation methods.
In particular, recent results using hybrid architectures, wherein different hardware is used to simulate the bosonic and fermionic degrees of freedom~\cite{Gonzalez_Cuadra:2022a,Zache:2023a}, allow for significant improvements in the simulation of lattice gauge theories.
For example, in such an encoding scheme, the implementation of any unitary in the plaquette term becomes trivial, having zero non-Clifford cost.
This may yield 5 -- 6 orders of magnitude improvement for SU(3) in the cost of qubitized simulation relative to the numbers we have published.

Two other related points of interest in the simulation of lattice gauge theories is the choice of model and the topic of gauge invariance.
In terms of considering different physical models, one natural direction to explore is investigating the peculiarities of mapping various lattice fermions (Wilson~\cite{Wilson:1974sk}, overlap~\cite{Neuberger:1997fp}, domain wall~\cite{Kaplan:1992bt}, etc.) onto qubits.
More broadly, as numerous formulations of LGTs emerge (purely bosonic~\cite{zohar2018eliminating,zohar2019removing,pardo2023resource}, prepotential~\cite{zohar2013cold,Davoudi:2022xmb}, loop-string-hadron~\cite{raychowdhury2020loop,davoudi2021search,Davoudi:2022xmb}, orbifold~\cite{Buser:2021a}, etc.), investigating their compatibility with state-of-the-art simulation techniques appears to be a crucial direction of research from the algorithmic perspective.

Regarding gauge invariance in the simulation, the qubitized algorithm presented in this work can break gauge invariance up to an error included in the $\epsilon$ simulation error of the dynamics.
If gauge invariance breaking is absolutely intolerable, then there are a few potential approaches to addressing it.
Firstly, one may choose to use discretized gauge theory approaches wherein simulation can be done without incurring gauge invariance breaking~\cite{Gustafson:2022a,Gustafson:2023a,Gustafson:2024a,Lamm:2024a}.
Secondly, one may choose to use specially adapted polynomials for quantum signal processing in the context of qubitized simulation which preserve gauge invariance, an open research problem.
Thirdly, one may forego addressing this question at the level of the logical algorithm, and instead address it in within the context of the error-correcting code that a prospective fault-tolerant device would make use of~\cite{weibe2024}.
Lastly, one may use an approach like the discrete gauge group Trotterization ~\cite{Gustafson:2022a,Gustafson:2023a,Gustafson:2024a,Lamm:2024a} which has built-in gauge invariance under the discrete group.

\section*{Acknowledgements}

This material is based upon work supported by the U.S.\ Department of Energy, Office of Science, National Quantum Information Science Research Centers, Quantum Systems Accelerator. SP was supported by the National Nuclear Security Administration's Advanced Simulation and Computing Program.

The authors benefited from a number of useful discussions with colleagues, whom we would like to acknowledge (in alphabetical order): Andrew Baczewski, Riley Chien, Andrew Landahl, Benjamin Morrison, and Stefan Seritan.

Sandia National Laboratories is a multimission laboratory managed and
operated by National Technology and Engineering Solutions of Sandia, LLC., a
wholly owned subsidiary of Honeywell International, Inc., for the U.S.\
Department of Energy's National Nuclear Security Administration under
contract DE-NA-0003525.

This paper describes objective technical results and analysis. Any
subjective views or opinions that might be expressed in the paper do not
necessarily represent the views of the U.S.\ Department of Energy or the
United States Government.

\bibliographystyle{apsrev4-1}
\bibliography{main}


\appendix

\section{The notion of Qubitization\label{app:qubitization}}

In the quantum simulation literature, the notion of \emph{Qubitization} has been repeatedly used for several different concepts.
To avoid potential confusion, we discuss those below.
\begin{enumerate}
    \item In the original work~\cite{Low:2019a}, \emph{Qubitization} referred to the process of constructing a special form of block encoding, the \emph{Szegedy walk operator} $W_H$, from
    a general block encoding $U_H$.
    The action of $U_H$ on a state of the form $\sket{0^m}\sket{\lambda}$, where $\sket{\lambda}$ is an eigenvector of $H$ with eigenvalue $\lambda$, can be written as
\begin{equation}
    U_H \sket{0^m}_a\sket{\lambda}_s
    = 
        \lambda \sket{0^m}_a \sket{\lambda}_s + \sqrt{1-\lambda^2} \sket{\bot^\lambda}_{as}\,,
\end{equation}
where the subscripts $s$ and $a$ indicate the system and ancillary qubit registers.
As compared to $U_H$, the quantum walk operator $W_H$ has an additional property that for each $\lambda$ the vector $\sket{\bot^\lambda}_{as}$ factorizes as $\sket{\bot^\lambda}_{as}=\sket{\bot^\lambda}_{a}\sket{\lambda}$:
\begin{equation}
    \label{eq:Wdeforig}
    W_H \sket{0^m}_a\sket{\lambda}_s
    = 
        \bigl(\lambda \sket{0^m}_a - \sqrt{1-\lambda^2} \sket{\bot^\lambda}_{a}\bigr)\sket{\lambda}_s\,,
\end{equation}
where the minus sign between the two terms is conventional and was added to matches the definition used in Ref.~\cite{Low:2019a}.

    The form of Eq.~\eqref{eq:Wdeforig} allows one to express its consecutive applications to the eigenstates of $H$ in terms of Chebyshev polynomials, and to readily utilize the machinery of Quantum Signal Processing (QSP)~\cite{low2017optimal} for implementing polynomial functions of $H$ using controlled calls to $W_H$.

    \item \emph{Qubitization} is also frequently used to term the entire procedure of Ref.~\cite{Low:2019a} for simulating dynamics governed by time-independent Hamiltonians which relies on QSP and controlled calls to~$W_H$.

    It is, however, important to note that simulating time evolution is not the only use case of $W_H$ in quantum simulation.
    For example, Szedegy walk operator is utilized in a highly-efficient implementation of Quantum Phase Estimation from Ref.~\cite{babbush2018encoding}.
    Such a construction is possible due to the fact that the spectrum of $W_H$ contains eigenvalues which are simple functions of the Hamiltonian eigenvalues:
    \begin{equation}
        \spec W_H \supset \{e^{\pm i \arccos \lambda}\}\,.
    \end{equation}

    3. \emph{Qubitization} is sometimes used as an umbrella term for all post-Trotter methods relying on block encoding.
    This is largely due to the fact that the approach of Ref.~\cite{Low:2019a} provides the most asymptotically efficient way simulating time-independent Hamiltonians, with complexity $\tilde{\mathcal{O}}(\|H\|t+\log1/\epsilon)$.

    Such terminology can be quite confusing, since near-optimal methods may use neither $W_H$ nor QSP.
    For example, one can simulate time evolution with complexity $\tilde{\mathcal{O}}(\|H\|t+\log1/\epsilon)$ by implementing the real and imaginary parts of the time evolution operator via Quantum Singular Value Transformation (QSVT), which are then added up with the Linear Combination of Unitaries~\cite{Gily_n_2019,Lin:2022a}.
    While QSVT is closely related to QSP, in this approach the construction of $W_H$ is not required.

    4. The notion of \emph{Qubitization} has also been used to denote the process of mapping physical degrees of freedom in discretized formulations of quantum field theory onto qubits~\cite{Huang:2021pwq,Alexandru:2022son}.
    This arguably most confusing convention is not related in any way to 1-3 above.

\end{enumerate}

    The authors did not agree on whether the usage of the term \emph{Qubitization} is permissible in the manuscript, as the truncated Dyson series algorithm~\cite{Low:2019b, Berry2020timedependent} is neither related to QSP nor does it employ the Szegedy walk operator.
    Similarly, the notion of ``Trotter'' decomposition is used throughout the paper as an umbrella term for methods based on product formul\ae.
    The decisions were made by vote.

\section{\label{app:lr} Bounding the Lieb-Robinson velocity}

Here we establish an upper bound on the Lieb-Robinson velocity $v_{LR}$ which is needed to lower bound the linear size of the HHKL block for the simulation. To accomplish this we employ the algorithm of Ref.~\cite{Wang:2020a} which establishes the tightest known Lieb-Robinson bounds for strictly local systems.

The algorithm begins by constructing the \emph{commutativity graph} $G$ of the Hamiltonian, in which vertices of $G$ correspond to terms of the Hamiltonian, and two vertices, $v_1$ and $v_2$, are connected by an edge if $[v_1,v_2]\neq 0$. Then one can write the associated Green's function of the commutativity graph, $G_{ij}(t)$, which is a solution to the differential equation
\begin{equation}
    \dot{G}_{ij}(t)=\sum_{k:\langle ik\rangle\in G}H_{ik}G_{kj}(t)\,,
\end{equation}
where $H_{ij}=2\sqrt{|h_i||h_j|}\delta(\langle ij\rangle\in G)$ and $h_i$ are the coefficients of the Hamiltonian term at site $i$. While a full solution is needed to bound the norm of the commutator for observables over disjoint regions, i.e. to obtain a Lieb-Robinson bound, the simpler task of obtaining the Lieb-Robinson velocity $v_{LR}$ only requires the matrix $H$. In particular, assuming translation invariance, we can work in the Fourier integral representation of the Green's function $G_{ij}(t)\coloneqq G_{I\alpha;J\beta}(t)$, defined by
\begin{equation}
    G_{ij}(t)=\int_{-\pi}^\pi \frac{d^dk}{(2\pi)^d} G_{\alpha\beta}^{(\vec{k})}(t) e^{i\vec{k}\cdot(\vec{r}_I-\vec{r}_J)}\,,
\end{equation}
where $G_{\alpha\beta}^{(\vec{k})}(t)$ is a solution to the differential equation
\begin{equation}
    \dot{G}_{\alpha\beta}^{(\vec{k})}(t)=\sum_{\gamma=1}^l H_{\alpha\gamma}^{(\vec{k})}G_{\gamma\beta}^{(\vec{k})}(t)\,.
\end{equation}
Here, $I,J\in [1,l]$ label unit cells of the lattice and $H_{\alpha\beta}^{(\vec{k})}$ is the Fourier transform of $H_{I\alpha;J\beta}$,
\begin{equation}
    H_{\alpha\beta}^{(\vec{k})}=\sum_J H_{I\alpha;J\beta}e^{-i \vec{k}\cdot(\vec{r}_I-\vec{r}_J)}.
\end{equation}

The task at hand is thus to construct $H_{\alpha\beta}^{(\vec{k})}$, from which we can bound the Lieb-Robinson velocity using its eigenvalues. We now describe a simple approach to accomplishing this task whenever the strictly local Hamiltonian can be represented entirely in terms of Pauli operators or Majorana fermion operators, a structure present in many physical Hamiltonians of interest.

One might wonder how this approach will be possible for the Kogut-Susskind Hamiltonian, given that the gauge operators do not evoke a Pauli or Majorana representation. According to Ref.~\cite[Lemma 13]{Tong:2022a}, the on-site electric field terms cannot change the Lieb-Robinson velocity, so they can be excluded from the commutativity graph. This allows us to additionally exclude the magnetic field terms from the commutativity graph, as the gauge operators only fail to commute with the electric field terms. The gauge operators will however remain in the gauge-matter interaction as a result of the fermionic anti-commutation relations between terms, but the gauge operators will only contribute a constant prefactor from their norm, so we can write the commutativity graph purely in terms of fermionic operators. We can now construct $H^{(\vec{k})}$ for our strictly local Hamiltonian expressed only as Pauli operators or Majorana fermions.

To do so, we note that we can encode any Pauli or Majorana operator on $n$ qubits as a binary string of length $2n$, namely, a vector $v\in\mathbb{F}_2^{2n}$. The first $n$ elements of $v$ index the presence of an $X$ or $\gamma$ operator, and the second $n$ elements indicate the presence of a $Z$ or $\bar{\gamma}$ operator, depending on the representation. Now consider the binary field extension $\mathbb{F}_2(x,x^{-1})^{2n}$ such that for vectors $v,w\in\mathbb{F}_2^{2n}$, we have $v+xw\in \mathbb{F}_2(x,x^{-1})^{2n}$. The variables $x$ and $x^{-1}$ allow us to represent translations of unit cells in the lattice, so for higher dimensions, additional extension variables, and their inverses, must be included.

The associated commutation relations of this representation are determined by $\sigma^\dagger \Lambda\sigma\mod 2\in\mathbb{F}_2(x,x^{-1})^{2n}$ where $\sigma$ is a $2\times n$ matrix composed of the binary representation vectors for the $n$ terms in a unit cell of the lattice and
\begin{equation}
    \Lambda=\unit\text{  (Majorana)  }\quad\text{and }\quad \Lambda=\begin{pmatrix}
    0 & \unit \\
    \unit & 0       
    \end{pmatrix}\text{  (Pauli).} 
\end{equation}
Also, in this representation, $\dagger$ corresponds to taking a transpose of the matrix and replacing $x$ with $x^{-1}$, or vice versa.

The next step is to define a function $f^{(\vec{k})}:\mathbb{F}_2(x,x^{-1})^{2n}\to \mathbb{C}^{2n}$ which maps $x\mapsto e^{ik}$, or in general $x^{\ell}\mapsto e^{ik\ell}$. We can then evaluate $f^{(\vec{k})}(\sigma^\dagger \Lambda \sigma\mod 2)$ for a given wavevector $k$. The last piece needed to construct the coefficient matrix $H^{(\vec{k})}$ is a vector storing the coefficients of the Hamiltonian terms. In particular, let $\vec{c}=\sqrt{2}(\sqrt{h_1},\sqrt{h_2},\ldots,\sqrt{h_n})$. Then, combining all of this together we have
\begin{equation}
    H^{(\vec{k})}=(\vec{c}\otimes \vec{c}^{\:\intercal})\odot f^{(\vec{k})}(\sigma^\dagger \Lambda \sigma\!\!\mod 2)\,,
    \label{eq:coefficient_matrix}
\end{equation}
where $\odot$ indicates elementwise multiplication.

At this point, we are ready to bound the Lieb-Robinson velocity, $v_{LR}$. According to Ref.~\cite{Wang:2020a}, we have
\begin{equation}
    v_{LR}\leq \min_{\kappa_0> 0}\frac{\omega(i\vec{\kappa}_0)}{\kappa_0}\,,
    \label{eq:LR_velocity}
\end{equation}
where $\omega(\vec{k})$ is the largest eigenvalue of $H^{(\vec{k})}$ and $\vec{\kappa}_0=\text{sgn}(\vec{r})\kappa_0$. Finally, $\vec{r}$ varies over all possible sign configurations for unit translations of the unit cell.

We now use this technique to bound the Lieb-Robinson velocity for the U(1) LGT in $d=2$. We use the Majorana represenation for simplicity, in which our fermionic terms of the Kogut-Susskind Hamiltonian become
\begin{align}
    H_M&=\frac{g_M}{2}\sum_x (\unit+i\gamma_x\bar{\gamma}_x)\,, \\
    H_{GM}&=\frac{g_{GM}}{4}\sum_{x,\hat{n}}\gamma_x(U-U^\dagger)\gamma_{x+\hat{n}}+i\gamma_x(U+U^\dagger)\bar{\gamma}_{x+\hat{n}}-i\bar{\gamma}_x(U+U^\dagger)\gamma_{x+\hat{n}}+\bar{\gamma}_x(U-U^\dagger)\bar{\gamma}_{x+\hat{n}}\,.
\end{align}
We construct our binary representation vectors in the field extension $\mathbb{F}_{2}(x,x^{-1},y,y^{-1})^{2n}$, which allows us to construct
\begin{equation}
    \sigma=\begin{pmatrix}
        1 & 1+x & 1 & x & 0 & 1+y & 1 & y & 0 \\
        1 & 0 & x & 1 & 1+x & 0 & y & 1 & 1+y
    \end{pmatrix}.
\end{equation}
Using the coefficients $g_M/2$ and $g_{GM}/4$ we can construct the coefficient vector $\vec{c}$ and employ Eq.~\eqref{eq:coefficient_matrix} to obtain $H^{(\vec{k})}$. Finally, we compute the largest eigenvalue and map $\vec{k}=(k_x,k_y)\mapsto i\kappa_0$. For the U(1) LGT in $d=2$, $\text{sgn}(\vec{r})$ yields the same eigenvalue for all possible translations of the unit cell.
Then employing Eq.~\eqref{eq:LR_velocity}, we minimize the eigenvalue over $\kappa_0$ where we have fixed the parameters $g_M=m$ and $g_{GM}=1/2a$ and used the parameter values $m=10$ and $a=0.1$ reported in Ref.~\cite{kan2021}. This results in
\begin{equation}
    v_{LR}\leq 53\,.
\end{equation}

\end{document}